\documentclass[12pt]{article}
\usepackage{epsfig,psfrag,cite}
\usepackage{amssymb,amsmath,psfrag}

\catcode`\@=11

\def \bl  {\begin{align*}}
\def \el  {\end{align*}}

\def \be  {\begin{equation}}
\def \ee  {\end{equation}}
\def \ba  {\begin{eqnarray}}
\def \ea  {\end{eqnarray}}
\def \baa {\begin{eqnarray*}}
\def \eaa {\end{eqnarray*}}
\def \bb  {\begin {thebibliography} }
\def \eb  {\end{thebibliography}}
\def \lab #1 {\label{#1}}

\newcommand\cA{\mathcal{A}}
\newcommand\cB{\mathcal{B}}
\newcommand\cU{\mathcal{U}}
\newcommand\cV{\mathcal{V}}
\newcommand\cW{\mathcal{W}}
\newcommand\cZ{\mathcal{Z}}

\newcommand\cN{\mathcal{N}}
\newcommand\C {\mathbb{C }}
\newcommand\CP {\mathbb{CP}}

\newcommand\im{\mathrm{i}}
\newcommand\la{\langle}
\newcommand\ra{\rangle}

\newcommand\delbar{\bar{\partial}}

\begin{document}

\thispagestyle{empty}

\begin{center}
\vskip 0.2truecm {\Large\bf
{\Large Dual Superconformal Invariance, Momentum Twistors and Grassmannians}
}\\
\vskip 1truecm
{\bf Lionel Mason$^{*}$ \& David Skinner$^{*,\dagger}$ \\
}

\vskip 0.4truecm
$^{*}${\it Mathematical Institute,\\ 
24-29 St. Giles', Oxford, OX1 3LB,\\ 
United Kingdom}\\

\vskip .2truecm 
$^{\dagger}${\it Perimeter Institute for Theoretical Physics,\\ 
31 Caroline St., Waterloo, ON, N2L 2Y5,\\ Canada}\\

\end{center}

\vskip 1truecm 
\centerline{\bf Abstract} 

\medskip

Dual superconformal invariance has recently emerged as a hidden
symmetry of planar scattering amplitudes in $\mathcal{N}=4$ super Yang-Mills theory.
This symmetry can be made manifest by expressing amplitudes in terms of `momentum twistors', as opposed to the usual twistors that make the ordinary superconformal properties manifest. The relation between momentum twistors and on-shell momenta is algebraic, so the translation procedure does not rely on any choice of space-time signature. We show that tree amplitudes and box coefficients are succinctly generated by integration of holomorphic $\delta$-functions in momentum twistors over cycles in a Grassmannian.  This is analogous to, although distinct from, recent results obtained by Arkani-Hamed {\it et al.} in ordinary twistor space.  We also make contact with Hodges' polyhedral representation of NMHV amplitudes in momentum twistor space.

\newpage
\setcounter{page}{1}\setcounter{footnote}{0}

\tableofcontents


\section{Introduction}
\label{sec:intro}

Dual superconformal symmetry has emerged as a powerful tool in the study of planar $\cN=4$ super Yang-Mills theory, providing stringent constraints on the structure of the scattering amplitudes~\cite{Drummond:2008vq}. This symmetry group has the same PSU$(2,2|4)$ structure as the standard superconformal group, but acts on `region momentum space' (differences between two region
momenta are the momenta of the particles in the scattering process), as opposed to ordinary Minkowski space. It is a hidden symmetry that may be viewed~\cite{Drummond:2009fd} as part of the Yangian structure of $\mathcal{N}=4$ SYM~\cite{Dolan:2003uh} predicted by AdS/CFT~\cite{Bena:2003wd}.

The MHV and six particle NMHV tree amplitudes were shown to be covariant under dual superconformal transformations in~\cite{Drummond:2008vq} (the MHV prefactor having weight under dual special conformal transformations) and this was generalised to all tree amplitudes in~\cite{Brandhuber:2008pf}. A systematic exploitation of both dual superconformal invariance and the BCFW recursion relations~\cite{Britto:2004ap,Britto:2005fq} then led to an explicit solution~\cite{Drummond:2008cr} for all $\mathcal{N}=4$ SYM trees. Although planar loop amplitudes are not thought to be directly invariant, the anomalous variation has a well-defined form~\cite{Drummond:2008vq,Drummond:2007cf}, so dual conformal invariance also imposes tight restrictions on the structure of loop amplitudes. In particular, it is consistent with the BDS conjecture~\cite{Bern:2005iz} (believed to be valid for four and five particles). Exact dual superconformal invariance has been explicitly confirmed for the coefficients of 1-loop box coefficients~\cite{Brandhuber:2008pf} (see also~\cite{Drummond:2008bq,Hall:2009xg}), while the anomalous dual conformal (though not dual \emph{super}conformal) symmetry has been verified for the 1-loop NMHV amplitudes themselves in~\cite{Brandhuber:2009xz,Elvang:2009ya} and extended to all amplitudes in~\cite{Brandhuber:2009kh}. It has also been observed at strong coupling via the work of Alday \& Maldacena~\cite{Alday:2007hr,Alday:2007he,Alday:2009yn}, where the dual conformal symmetry is identified as the isometry group of a copy of AdS$_5$ that is T-dual to the usual one~\cite{Berkovits:2008ic,Beisert:2008iq}.
 
In a somewhat separate line, the properties of tree amplitudes under
usual superconformal transformations can be made manifest by
transforming them into twistor
space~\cite{Hodges:2006tw,Hodges:2005aj,Hodges:2005bf,Mason:2009sa,ArkaniHamed:2009si},
where they reveal a rich geometric
structure~\cite{Mason:2009sa,Korchemsky:2009jv}. Building on this
work, Arkani-Hamed {\it et al.} have recently
shown~\cite{ArkaniHamed:2009dn} that $n$-particle N$^k$MHV tree
amplitudes can be obtained from a contour integral over a certain
Grassmannian manifold, where the integrand is naturally a function of
$n$ twistors in addition to the Grassmannian parameters. 
Furthermore,
different contours in the same Grassmannian lead to 1-loop box
coefficients.

\bigskip

Now, just as twistor space can be introduced as the spin space of the
ordinary superconformal group -- twistors being in the fundamental
representation of PSU$(2,2|4)$ -- we can also introduce a twistor
space for the \emph{dual} superconformal group. Following Hodges, who
first introduced them in~\cite{Hodges:2009hk}, we will refer to these
dual superconformal twistors as \emph{momentum twistors}, as they
relate to (region) momentum space in essentially the same way that
ordinary twistors relate to space-time. As a result, momentum space
amplitudes may be transformed to momentum twistor space by a purely
algebraic procedure. In particular, \emph{no choice of space-time
  signature is implied} and many of the problems of the usual twistor
approach are thereby avoided. 

In this paper, we show that amplitudes and box coefficients may also
be generated from a contour integral over cycles in a Grassmannian
that depend on momentum twistors. Thus, as well as the dihedral
symmetry, \emph{dual} superconformal invariance is now made
manifest. The momentum twistor and original twistor generating
functions are strikingly similar. However, for N$^k$MHV amplitudes,
the Grassmannian we deal with is G$(k,n)$ -- the parameter space of
$k$-planes in $\mathbb{C}^n$ -- rather than G$(k+2,n)$ and,
correspondingly, our integrand involves determinants of $k\times k$
matrices rather than $(k+2)\times(k+2)$ matrices. Explicit
computations are thus considerably easier than
in~\cite{ArkaniHamed:2009dn}.  We also share many of the benefits of
their framework.  These formulations make it easy to see the dihedral
symmetry and parity invariance of the amplitudes.  Furthermore, highly
non-trivial identities in momentum space -- such as the equivalence of
different BCFW decompositions of a tree amplitude, or the IR equations
that enforce vanishing of certain combinations of box
coefficients~\cite{Roiban:2004ix} -- become simple applications of the global
residue theorem (a higher-dimensional generalisation of Cauchy's
theorem, see {\it e.g.}~\cite{GH,T}). Their formulation makes the
usual superconformal invariance explicit, whereas ours makes the dual
superconformal invariance explicit.

Writing amplitudes in the momentum twistor representation has several
benefits. Most obviously, dual superconformal invariance is made
manifest. Secondly, the various terms that contribute to an amplitude
or box coefficient ({\it e.g.} $R_{n;ab}R_{n;cd}$,
$R_{n;ab}R_{n;ab;cd}$, $R_{n;ab}R_{n;ab;ad}^{ba}$ and
$R_{n;ab}R_{n;bd}^{ab}$ for the N$^2$MHV tree~\cite{Drummond:2008cr})
are all placed on equal footing. Third, as in usual twistor space, the
global residue theorem provides the key mathematical machinery
underlying the non-trivial relations between different \emph{sums} of
invariants. Finally, the momentum twistor representation gives such
dual superconformal invariants a \emph{geometric} meaning. This is
most transparent for the basic invariants $R_{n;ab}$ -- revealed in
section~\ref{sec:Rint} as the condition that five momentum
supertwistors be linearly dependent -- but more complicated dual
superconformal invariants are also intimately connected with the
geometry of the Grassmannian. 

\bigskip

The plan of the paper is as follows. In section~\ref{sec:dcimomtwist},
we explain how to transform $\mathcal{N}=4$ superamplitudes to
momentum twistor space.  This fills the gap in~\cite{Hodges:2009hk},
where the momentum twistor form of the \emph{super}conformal
$R_{n;ab}$ was only posited. In section~\ref{sec:Grassmannian} we
introduce the Grassmannian proposal, motivating it as a natural
generalisation of the expression for $A^{\rm NMHV}_{5,0}/ A^{\rm
  MHV}_{5,0}$ found in section~\ref{sec:Rint}. In
section~\ref{sec:contours} we give a further set of examples of the
use of the Grassmannian to generate amplitudes and box
coefficients. We find that the specific contours needed to isolate a
given tree amplitude or box coefficient closely correspond to the
contour specifications made in~\cite{ArkaniHamed:2009dn} (curiously,
they differ only by the simple cyclic shift $i\to i\!+\!1$ in the external
states). In particular, after finding contours that yield
superconformal invariants which do not correspond to any object at
tree level or one loop, Arkani-Hamed {\it et
  al.}~\cite{ArkaniHamed:2009dn} made the bold conjecture that their
Grassmannian generating function really computes the leading
singularity~\cite{Buchbinder:2005wp,Cachazo:2008dx,Cachazo:2008vp,
  Cachazo:2008hp,Spradlin:2008uu} of amplitudes to \emph{all
  loops}. Since every contour choice made in~\cite{ArkaniHamed:2009dn}
also has a meaning here, we also make the analogous conjecture in this
dual superconformal context.

As mentioned above, momentum twistors were originally introduced by
Hodges~\cite{Hodges:2009hk} where he interpreted NMHV tree amplitudes
as supersymmetric volumes of certain dihedrally symmetric polytopes in
dual twistor space. In section~\ref{sec:polytopes} we give a formal
argument to show how such polytopes are related to our Grassmannian
formulation by Fourier transform. We conclude in section~\ref{sec:conclusions}.

\emph{Note added:} immediately after this paper appeared on the arXiv, a direct relationship between the Grassmannian formul\ae\ in twistor space and momentum twistor space was derived in~\cite{ArkaniHamed:2009vw}.


\section{Momentum Twistors and Dual Conformal Invariance}
\label{sec:dcimomtwist}

An on-shell $\cN=4$ supermultiplet
\be
	\Phi(\lambda,\tilde\lambda,\eta) = G^+(\lambda,\tilde\lambda) + \eta^a\Gamma_a(\lambda,\tilde\lambda) + \cdots
	+\frac{\epsilon_{abcd}}{4!}\eta^a\eta^b\eta^c\eta^dG^-(\lambda,\tilde\lambda)\ .
\label{supermultiplet}
\ee
depends on bosonic spinor momenta\footnote{$A=0,1$ and $A'=0',1'$ are anti self-dual and self-dual Weyl spinor indices, while $a=1,\dots,4$ is an $R$-symmetry index. We typically suppress these indices in what follows.} $(\lambda_A, \tilde\lambda_{A'})$ and a fermionic variable $\eta^a$ that counts the helicity of the component fields. In the planar sector of $n$-particle scattering amplitudes, the colour-ordering allows us to naturally encode the $n$ such supermomenta into $n$ region supermomenta $(x_i,\theta_i)$, defined up to translation by (see figure~\ref{fig:region})
\be
	x_i - x_{i+1}\equiv\lambda_i\tilde\lambda_i \qquad \theta_i- \theta_{i+1}\equiv\lambda_i\eta_i\ .
\label{regionmom}
\ee 
In this paper, we will be primarily interested not in the usual superconformal group acting on space-time, but rather the \emph{dual} superconformal group. This acts on the region momenta $(x_i,\theta_i)$ in exactly the same manner as the usual superconformal group acts on space-time~\cite{Drummond:2008vq}. It is thus possible to construct a twistor space for the region momenta. To prevent confusion with the usual twistor space and its dual, we follow~\cite{Hodges:2009hk} in calling this space \emph{momentum twistor space} -- it is the space of the fundamental representation of the \emph{dual} conformal group. 

Working with $(x,\theta)$ ensures that the supermomentum constraints
\be
	\sum_i p_i =0\qquad \sum_i \lambda_i\eta_i =0
\ee
are automatically satisfied. The picture is of a polygon in the space
of region momenta, all of whose edges are null ray segments
corresponding to the supermomenta of the external particles. Inspired
by the AdS/CFT correspondence~\cite{Alday:2007hr} it was shown
in~\cite{Drummond:2007aua,Brandhuber:2007yx} that the expectation
value of a Wilson 
loop stretched along this polygonal contour reproduces the MHV
amplitude\footnote{So far, the correspondence has been checked for $n$
particles at 1-loop~\cite{Brandhuber:2007yx} and up to 6 particles at
2-loops~\cite{Drummond:2008aq,Bern:2008ap}. (The $n$ particle 2-loop
MHV amplitude has recently been computed in~\cite{Vergu:2009tu}, where
the answer is expressed as a sum of conformal integrals, while the
corresponding Wilson loops computation was performed numerically
in~\cite{Anastasiou:2009kna}. Differences in methodology mean that
these results have not yet been compared.)}. From the point of view of
the Wilson loop, the dual superconformal symmetry of the amplitude is
just the usual superconformal symmetry. Thus, one may equivalently
think of momentum twistor space as the standard twistor space
associated to this Wilson loop. 


\subsection{Basics of twistor geometry}
\label{sec:basics}

In this subsection we briefly review the basics of twistor geometry and its correspondence with space-time. This correspondence can equally be viewed as being between usual twistor space and (conformally compactified) space-time, or between momentum twistor space and the (dual conformally compactified) space of region momenta. Most of the material  here is readily available in standard twistor theory texts (see {\it e.g.}~\cite{Penrose:1986ca,Penrose:1972ia,Ward:1990vs,Huggett:1986fs}), but since many of the following formul\ae\ are useful in translating the dual superconformal invariants into momentum twistor space, we include their derivation to make this paper self-contained. Readers for whom the twistor correspondence is familiar (or those willing to take our translation to momentum twistor space on trust) may prefer to skip ahead to section~\ref{sec:translation}.

\bigskip

Conformally compactified Minkowski space may be described as the SO$(2,4)$-invariant (Klein) quadric
\be
	T^2+V^2-W^2-X^2-Y^2-Z^2=0\ ,
\label{KleinQ1}
\ee
where $\{T,V,W,X,Y,Z\}$ are homogeneous coordinates for $\mathbb{RP}^5$. It is convenient to package these six coordinates into $X_{\alpha\beta}=-X_{\beta\alpha}$ (where $\alpha,\beta,\ldots = 0,1,2,3$) as follows
\be
\begin{aligned}
	X_{01} &=W-V\,, &\quad& X_{02}=\frac{1}{\sqrt{2}}(Y+\im X)\,, &\quad& X_{03}=\frac{\im}{\sqrt{2}}(T-Z)\,,\\
	X_{12} &=-\frac{\im}{\sqrt{2}}(T+Z)\,, &\quad& X_{13}= \frac{1}{\sqrt{2}}(Y-\im X)\,, &\quad& X_{23}=\frac{1}{2}(V+W)\ .
\end{aligned}
\ee
whereupon the quadric~\eqref{KleinQ1} becomes\footnote{The totally skew tensor $\epsilon^{\alpha\beta\gamma\delta}=\epsilon^{[\alpha\beta\gamma\delta]}$ (with $\epsilon^{0123}=+1$) is a canonical structure for the conformal group SO$(2,4)\simeq{\rm SU}(2,2)$.}  
\be
	\epsilon^{\alpha\beta\gamma\delta}X_{\alpha\beta}X_{\gamma\delta} = 0\ .
\label{KleinQ}
\ee
This condition turns out to be equivalent to the simplicity constraint $X_{\alpha[\beta}X_{\gamma\delta]}=0$, so an arbitrary skew $X$ satisfies~\eqref{KleinQ} -- and thus corresponds to a point in space-time -- if and only if
\be
	X_{\alpha\beta} = A_{[\alpha}B_{\beta]}
\ee
for some $A$, $B$ in \emph{twistor space}. It is often convenient to work with complexified space-time, with $X_{\alpha\beta}$ viewed as homogeneous coordinates on $\mathbb{CP}^5$. Then twistor space is a copy of $\mathbb{CP}^3$ and may be described by homogeneous coordinates\footnote{With Penrose conventions, the twistor space with coordinates $Z^\alpha$ would usually be taken as primary, and the $W_\alpha$ space referred to as `dual'. It is unfortunate that this clashes with the prevalent conventions in perturbative gauge theory, whereby MHV amplitudes involve unprimed/undotted spinors $|\lambda\ra$ and so live most naturally on Penrose's dual space. We will work with perturbative gauge theory conventions in this paper.} $[W_\alpha]$ subject to the equivalence relation $[W_\alpha]\sim [rW_\alpha]$ for any non-zero complex scaling $r$. The two (distinct) points $A,B$ determine a line\footnote{In the dual $\mathbb{CP}^3$ -- Penrose's twistor space -- the equations $A_\alpha Z^\alpha=0$ and $B_\alpha Z^\alpha=0$ each determine a plane ($\mathbb{CP}^2$), whose intersection is again a line ($\mathbb{CP}^1$).} in $\mathbb{CP}^3$, so points of conformally compactified, complexified space-time correspond to holomorphic lines $\mathbb{CP}^1\subset\mathbb{CP}^3$. Conversely, an arbitrary point $W\in\mathbb{CP}^3$ lies on the line $[A\wedge B]$ if and only if
\be
	X_{[\alpha\beta}W_{\gamma]}=0
\label{incidence}
\ee
so that $W_\alpha$ is a linear combination of $A_\alpha$ and $B_\alpha$.

\bigskip

\begin{figure}[h]
\begin{center}
	\includegraphics[height=45mm]{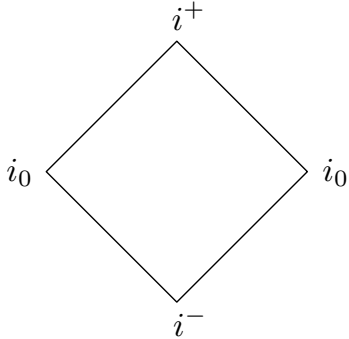}
\end{center}
\caption{{\it The Penrose diagram of Minkowski space. Spacelike infinity $i_0$ is a single point, and $i_0$ and $i^\pm$ are all identified in the conformal compactification. They thus correspond to the same point on the Klein quadric, and the same line $I$ in twistor space. The rest of the conformal boundary -- null infinity -- corresponds to points $X$ on the Klein quadric that obey $X^{\alpha\beta}I_{\alpha\beta}=0$, or twistor lines that intersect the distinguished line $I$.}}
\label{fig:Penrose}
\end{figure}

The Klein quadric has a natural conformal structure: two points $X$ and $Y$ on the Klein quadric ({\it i.e.,} conformally compactified space-time) are  \emph{null-separated} if and only if
\be
	\epsilon^{\alpha\beta\gamma\delta}X_{\alpha\beta}Y_{\gamma\delta}=0\ ,
\label{nullseparated}
\ee
in which case their associated twistor lines intersect. However, since $X_{\alpha\beta}$ and $Y_{\alpha\beta}$ are homogeneous coordinates, there is no natural scale for any non-zero value of $X^{\alpha\beta}Y_{\alpha\beta}$. To pick a preferred scale, breaking the conformal group to the Poincar{\' e} group, one introduces the fixed \emph{infinity twistor}, defined by
\be
	I_{\alpha\beta} \equiv	\begin{pmatrix}
						\epsilon^{A'B'} & 0\\
						0 & 0
					\end{pmatrix}\ ,
\label{infinitytwistor}
\ee
that represents a fixed point in conformally compactified space-time, taken to be the vertex of the `lightcone at infinity' (see figure~\ref{fig:Penrose}). The infinity twistor allows us to define a metric
\be
	(x-y)^2 \equiv 
	\frac{X^{\alpha\beta}Y_{\alpha\beta}}{I_{\gamma\delta}X^{\gamma\delta} I_{\rho\sigma}Y^{\rho\sigma}}
\label{metric}
\ee
that is independent of rescalings of the homogeneous coordinates. Thus, if $x$ and $y$ are represented by the twistor lines $[A\wedge B]$ and $[C\wedge D]$ respectively, their Minkowski separation is
\be
	(x-y)^2 = \frac{\epsilon(A,B,C,D)}{\la AB\ra\la CD\ra}
\label{twistdist}
\ee
in terms of twistor variables, where $\epsilon(A,B,C,D)\equiv \epsilon^{\alpha\beta\gamma\delta}A_\alpha B_\beta C_\gamma D_\delta$ and
\be 
	\la AB\ra \equiv I^{\alpha\beta}A_\alpha B_\beta
\ee
is the standard spinor inner product.

The infinity twistor also plays an important role as a projection operator. Raising indices with the $\epsilon$-symbol, the dual infinity twistor
\be
	I^{\alpha\beta} \equiv \frac{1}{2}\epsilon^{\alpha\beta\gamma\delta}I_{\gamma\delta} 
	=\begin{pmatrix} 0 & 0 \\ 0 & \epsilon^{AB} \end{pmatrix}
\label{dualinfinitytwistor}
\ee
projects $W_\alpha$ onto its secondary (unprimed) part
\be
	 I^{\alpha\beta}W_{\beta} = (0,\lambda^A)
\label{project1}
\ee
where $\lambda^A$ is a left-handed Weyl spinor. We will abuse notation
somewhat to informally write $\lambda^A = I^{\alpha\beta}W_{\beta}$. The remaining components of $W_\alpha$ behave as a spinor of opposite chirality and we often decompose twistors into their constituent spinors $W_\alpha=(\mu^{A'},\lambda_A)$. 
In these coordinates, the usual coordinates $x^{AA'}$ on affine space-time are obtained from $X_{\alpha\beta}$ by
\be
	X_{\alpha\beta} =	\begin{pmatrix}
						-\frac{1}{2}\epsilon^{A'B'}x^2 & -\im x^{A'}_{\ \ B}\\
						 \im x_A^{\ \ B'} & \epsilon_{AB}
					\end{pmatrix}
	\ ,\qquad
	X^{\alpha\beta} = 	\begin{pmatrix}
						\epsilon_{A'B'}& -\im x_{A'}^{\ \ B}  \\
						\im x^A_{\ \ B'}& -\frac{1}{2}\epsilon^{AB}x^2 
					\end{pmatrix}\ ,
\label{Xcomps}
\ee
where, if $X$ is the line through $U=(\mu_U,\lambda_U)$ and $V=(\mu_V,\lambda_V)$ in twistor space, then
\be
	x^{AC'} = \im \frac{\left(\mu_U\lambda_V-\mu_V\lambda_U\right)^{AC'}}{\la U\,V\ra} 
\ee
The incidence relation~\eqref{incidence} becomes
\be
	\mu^{A'}=-\im x^{AA'}\lambda_A
\label{affincidence}
\ee
(as many readers will find familiar), while the distinguished line corresponding to the infinity twistor is $\lambda_A=0$. Likewise, the displacement $x-y$ may be written in twistor variables as
\be
\begin{aligned}
	(x-y)^{CA'} &= I_{\alpha\beta}\, I^{\gamma\delta}\,
	\frac{\epsilon^\beta(\,\cdot\,,A,B,C)D_\delta
	- \epsilon^\beta(\,\cdot\,,A,B,D)C_\delta}{\la AB\ra\la CD\ra}\\
	& = I_{\alpha\beta}\, I^{\gamma\delta}\,
	\frac{\epsilon^\beta(\,\cdot\,,C,D,B)A_\delta
	- \epsilon^\beta(\,\cdot\,,C,D,A)B_\delta}{\la AB\ra\la CD\ra}\ .
\end{aligned}
\label{displacement}
\ee
Though not conformally invariant by themselves, equations~\eqref{twistdist} \&~\eqref{displacement} will be of particular use in translating (dual) conformal invariants such as $R_{n;ab}$ to (momentum) twistor space.

\bigskip

All of the above has a straightforward generalization to supertwistors. Conformally compactified chiral superspace may be viewed as the space of lines in $\CP^{3|4}$ (for $\cN=4$ supersymmetry), with homogeneous coordinates $[\cW_I] = [W_\alpha,\chi_a]$. A line through points $\cU$ and $\cV$ in supertwistor space is described by the simple (graded-)skew supertwistor $X_{IJ} = \cU_{[I}\cV_{J\}}$. The incidence relation~\eqref{incidence} generalizes to
$X_{[IJ}\cW_{K\}}=0$, or
\be
	\mu^{A'}=-\im x^{AA'}\lambda_A\,,\qquad \chi_a = \theta^A_{\ a}\lambda_A
\label{superincidence}
\ee
in the basis determined by the infinity twistor. Here, $(x,\theta)$ are coordinates on an affine patch of chiral superspace and are given in terms of the components of the supertwistors $\cU$, $\cV$ as
\be
	(x,\theta) = \left(\im \frac{\mu_V\lambda_U-\mu_U\lambda_V}{\la UV\ra}\ ,\ 
	\frac{\chi_V\lambda_U - \chi_U\lambda_V}{\la UV\ra}\right)\ .
\ee
Note also that
\be
	\theta^A_{\ r} = \frac{I^{\alpha\beta} (\cU_{[\beta}\cV_{r]})}{\la U\,V\ra}\ ,
\label{thetaproj1}
\ee
so that $\theta$ is simply the projection of the fermionic part of $[\cU\wedge\cV]$ using the infinity twistor as in~\eqref{dualinfinitytwistor}. This expression will be useful when translating the numerators of the dual superconformal invariants.

\begin{figure}[h]
\begin{center}
	\includegraphics[height=65mm]{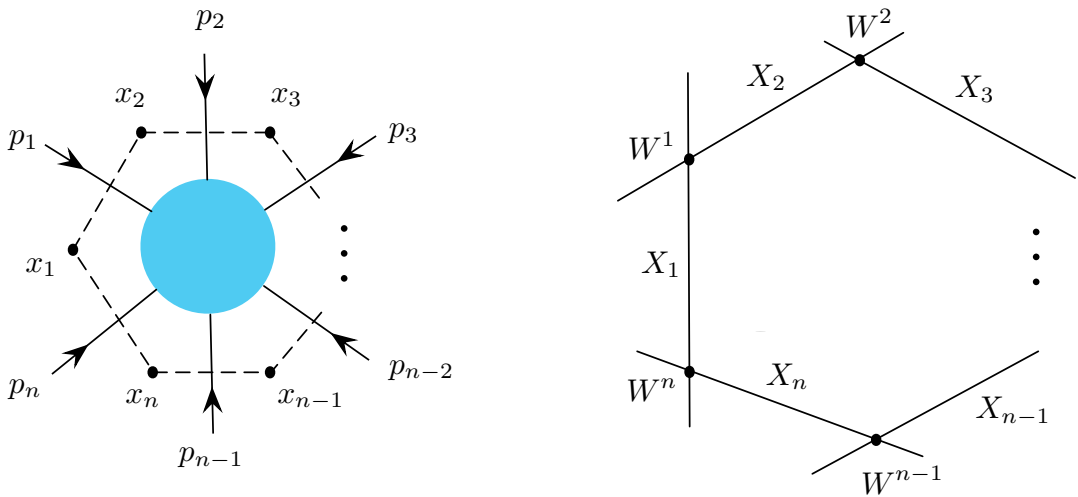}
\end{center}
\caption{{\it A scattering amplitude in momentum space, together with the corresponding array of (generically skew) intersecting lines in momentum twistor space. The diagram illustrates the labelling of region momenta $x_i$. Our conventions are such that $x_{ij} = \sum_{k=i}^{j-1} p_k$ and therefore $X_i\cap X_{i+1}=W^i$. Note that the array of twistor lines corresponds precisely to the polygonal contour of the Wilson loop in $x$-space, with edges and vertices interchanged.}}
\label{fig:region}
\end{figure}

So far, the geometric correspondence we have outlined holds equally for the usual twistor space of standard space-time and for the momentum twistor space associated to the region momenta. However, the cyclic ordering inherent in the definition of region momenta introduces some special features that we now discuss.

Null geodesics in space-time correspond to a unique twistor (up to overall scaling): given a point $x_0$ on the ray
\be
	x(t)= x_0+t\lambda\tilde\lambda\ ,
\ee
all the components of $W$ are fixed by the incidence relation~\eqref{affincidence} after identifying the unprimed part $I^{\alpha\beta}W_\alpha$ of $W$ with the unprimed spinor $\lambda^A$ determined by the ray. Thus, associated to our null polygon in region momentum space, there are $n$ twistors $W^i$ -- one for each external particle (or each edge of the Wilson loop). This determines a polygon in twistor space whose edges are the lines $X_i$ corresponding to the region momenta $x_i$.  Our conventions are that the region momentum $x_i$ corresponds to the line $X_i$ through points
$W^{i-1}$ and $W^{i}$ in momentum twistor space (see figure~\ref{fig:region}). Thus
\be
	(X_i)_{\alpha\beta}=W^{i-1}_{[\alpha}W^{i}_{\beta]}\qquad\hbox{or}\qquad
	(x_i)^{AC'} = \frac{I^{\alpha\beta}\left(W^{i-1}_\beta W^{i\, C'}-W^{i-1\,C'}W_\beta^i\right)}{\la i\!-\!1\ i\ra}
\ee 
With these conventions, $W_i = X_i\cap X_{i+1}$ so 
\be
	\mu^{iA'} = -\im x_i^{AA'}\lambda_A^i\qquad\hbox{and}\qquad 
	\mu^{iA'} =-\im x_{i+1}^{AA'}\lambda_A^i \qquad\Rightarrow\qquad (x_i-x_{i+1})^{AA'}\lambda_A^i=0\ .
\label{ixincidence}
\ee
For the fermionic components, we likewise have
\be
	(\theta_i)^{A}_{\ r} = \frac{I^{\alpha\beta} \left(\chi^{i-1}_r W^i_\beta-W^{i-1}_\beta\chi^i_r\right)}{\la i\!-\!1\ i\ra}\ ,
\label{thetaproj}
\ee
while the incidence relations read
\be
	\chi^i_{\ a} = \theta^{A}_{i\,a}\lambda^i_A\qquad\hbox{and}\qquad
	\chi^i_{\ a} = \theta^{A}_{i+1\,a}\lambda^i_A\qquad\Rightarrow\qquad (\theta_i-\theta_{i+1})^{AA'}\lambda_A^i =0\ .
\ee
Thus the incidence properties are consistent with the conventions $x_i-x_{i+1}=\lambda^i\tilde\lambda^i= p_i$ and $\theta_i-\theta_{i+1}= \lambda^i\eta^i$ of equation~\eqref{regionmom}. As a corollary, we can identify $\tilde\lambda^i$ and $\eta^i$ in terms of twistor variables as
\be
\begin{aligned}
	\tilde\lambda_i &=  -\im \frac{\mu^{i-1}\la i\ i\!+\!1\ra + \mu^i\la i\!+\!1\ i\!-\!1\ra + \mu^{i+1}\la i\!-\!1\ i\ra}
	{\la i\!-\!1\ i\ra\la i\ i\!+\!1\ra}\\
	\eta_i &= - \frac{\chi^{i-1}\la i\ i\!+\!1\ra + \chi^i\la i\!+\!1\ i\!-\!1\ra + \chi^{i+1}\la i\!-\!1\ i\ra}
	{\la i\!-\!1\ i\ra\la i\ i\!+\!1\ra}\ .
\end{aligned}
\ee
Together with the fact that the momentum spinor $\lambda_i$ is also
the secondary part of the twistor,
this allows us to express translate amplitudes from supermomenta to momentum twistors.   
However, it is clear from this formula that although there is one
twistor for each external particle, the relation between the 
momenta and the twistor is not localized -- to specify a particle's momentum, one needs 
knowledge of both its corresponding twistor and those of its nearest neighbours in the given colour-ordering.


\subsection{Translating the dual superconformal invariants}
\label{sec:translation}

The simplest dual superconformal invariants appearing in the
scattering amplitudes are
\be
R_{t;rs} \equiv \frac{\la r-1\, r\ra\la s-1\, s\ra\
  \delta^{0|4}\!\left(\Xi_{t;rs}\right)} {x_{rs}^2\la
  t|x_{ts}x_{sr}|r-1\ra\la t|x_{ts}x_{sr}|r\ra\la
  t|x_{tr}x_{rs}|s-1\ra\la t|x_{tr}x_{rs}|s\ra} 
\label{Rdef}
\ee where 
\be 
	\Xi_{t;rs} \equiv \la t|x_{ts}x_{sr}|\theta_{rt}\ra +
	\la t|x_{tr}x_{rs}|\theta_{st}\ra\ .
\ee 
Here, $x_{ij}\equiv x_i-x_j= \sum_{k=i}^{j-1} \lambda_k\tilde\lambda_k$ denotes a partial sum of
the external momenta, and $\theta_{ij}\equiv \theta_i-\theta_j$
likewise denotes the partial sum $\sum_{k=i}^{j-1} \lambda_k\eta_k$. 
In this subsection, we show that $R_{t;rs}$ has a simple expression in 
terms of momentum twistors. This provides the supersymmetric formula 
needed by Hodges in his interpretation of NMHV tree amplitudes
as volumes of certain polyhedra~\cite{Hodges:2009hk} (and assumed by him in equation~(30) of that paper). The formula we obtain has manifest dual conformal, but not dual \emph{super}conformal symmetry. A formula that has the full dual superconformal symmetry manifest will be obtained in the following subsection, and leads naturally to our Grassmannian generating function.

\bigskip

Consider first the denominator of~\eqref{Rdef}. It follows immediately from~\eqref{metric} and $X_r = [W^r\wedge W^{r-1}]$ that
\be
	x_{rs}^2 = \frac{\epsilon(r-1,r,s-1,s)}{\la r\!-\!1\ r\ra\la s\!-\!1\ s\ra}\ .
\ee
Likewise, the displacement formula~\eqref{displacement} for $x_{ts}$ and $x_{sr}$ gives
\be
	\la t|x_{ts} x_{sr}|r-1\ra = \frac{\epsilon(t,s-1,s,r-1)}{\la s\!-\!1\ s\ra}\ .
\ee
This allows us to translate the denominator. Exactly the same reasoning can be used for the numerator, where it is helpful to first rewrite $\Xi_{t;rs}$ using the identity
\be
	\la t|x_{ts}x_{sr}|\theta_{rt}\ra + \la t|x_{tr}x_{rs}|\theta_{st}\ra = 
	x^2_{rs}\la t|\theta_r\ra + \la t|x_{ts} x_{sr}|\theta_r\ra + \la t|x_{tr}x_{rs}|\theta_s\ra\ .
\ee
From equation~\eqref{thetaproj} we find
\be
	x^2_{rs}\la t|\theta_t\ra = \frac{\epsilon(r\!-\!1,r,s\!-\!1,s)}{\la r\!-\!1\ r\ra\la s\!-\!1\ s\ra}\chi^t
\label{fermitrans1}
\ee
and
\be
	\la t|x_{tr}x_{rs}|\theta_s\ra = \frac{\epsilon(t,r\!-\!1,r,s\!-\!1)\chi^s - \epsilon(t,r\!-\!1,r,s)\chi^{s-1}}
	{\la r\!-\!1\ r\ra\la s\!-\!1\ s\ra}
\label{fermitrans2}
\ee
Combining these and analogous terms shows that $R_{t;rs}$ is written in momentum twistor variables as
\be
	R_{t;rs} = \frac{\delta^{0|4}(\chi^t\epsilon(r-1,r,s-1,s) + \hbox{cyclic})}
			{\epsilon(t,r-1,r,s-1)\epsilon(r-1,r,s-1,s)\epsilon(r,s-1,s,t)\epsilon(s-1,s,t,r-1)\epsilon(s,t,r-1,r)}\ .
\label{Rmomtwist}
\ee
We will often denote these basic dual superconformal invariants -- the ratio of the 5-particle N$^1$MHV tree amplitude to the 5-particle MHV tree -- by $R_{1,5}$. When we wish to make explicit the dependence on particular external twistors, we will use the notation $R_{1,5}(r-1,r,s-1,s,t)$.

\bigskip

Equation \eqref{Rmomtwist} makes various properties of these dual superconformal invariants transparent. Firstly, we see that all the factors involving infinity twistors (spinor products) have cancelled out and~\eqref{Rmomtwist} is written purely in terms of skew products of momentum twistors, together with a fermionic $\delta^{0|4}$-function. Thus $R_{1,5}$ is \emph{manifestly} invariant under the maximal bosonic subgroup of the dual superconformal group. More importantly $R_{1,5}(a,b,c,d,e)$ makes clear that these invariants depend on \emph{five} external states, and is skew symmetric in its arguments. Thus, various identities of the form
\be
	R_{r+2;s,r+1} \equiv R_{r;r+2,s}
\ee
follow immediately, whereas they are somewhat obscure in the momentum space expression~\eqref{Rdef}. Such identities are useful, for example in proving the dual superconformal invariance of all one-loop NMHV box cofficients~\cite{Drummond:2008bq,Elvang:2009ya,Brandhuber:2009kh}. Equation~\eqref{Rmomtwist} also shows that $R_{1,5}$ becomes singular\footnote{Of course, a given singularity may be absent in some components of the $\mathcal{N}=4$ supermultiplet.} whenever any of the denominator factors $\epsilon(a,b,c,d)$ vanish, {\it i.e.} whenever any four of the five twistors become coplanar. These singularities are physical whenever $\{a,b,c,d\}$ form two cylically adjacent pairs, otherwise they are spurious and must cancel in the overall expression for the amplitude.

We also emphasize that, since the translation between the $(x,\lambda,\theta)$ and $\cW=(W,\chi)$ variables was purely algebraic, \emph{it does not rely on any choice of space-time signature}. This is standard for twistor constructions (such as the Penrose transform) that move between twistor space and space-time, but is in marked contrast to Witten's half-Fourier transform~\cite{Witten:2003nn} between on-shell momentum space and (standard) twistor space, that is well-defined only for real twistors and (2,2)-signature space-time. In the present case, \eqref{Rmomtwist} may be analytically continued as a holomorphic function of five momentum supertwistors, just as~\eqref{Rdef} may be analytically continued into complex momentum space.


\subsection{$R_{t;rs}$ as linear dependence of five momentum supertwistors}
\label{sec:Rint}

Although easily shown, \eqref{Rmomtwist} does not make the full \emph{super}conformal symmetry manifest. It is revealing to reformulate $R_{1,5}$ so as to bring this out. The reformulation will show that $R_{1,5}$ has a simple geometric meaning -- it is the condition that five supertwistors are linearly dependent. Furthermore, this reformulation provides the key to generalizing $R_{1,5}$ to other dual superconformal invariants needed for $n$-particle N$^k$MHV amplitudes.

We wish to prove that 
\be
	R_{1,5}=\int_{\mathbb{CP}^4} \frac{D^4T}{T^1\cdots T^5} \,\bar\delta^{4|4}\!\left(\sum_{i=1}^5 T_i\cW^i\right)\ .
\label{Rint}
\ee
Let us first explain the notation. The integral is to be taken over a copy of $\mathbb{CP}^4$ with homogeneous coordinates $[T_i]$ (where $i=1,\ldots,5$) and 
\be
	D^4T:=\frac{1}{5!}\epsilon^{ijklm}T_i dT_j\wedge dT_k\wedge dT_l\wedge dT_m
\label{projmeasure}
\ee
is the canonical top holomorphic form of homogeneity $+5$. The weight of this form is balanced by the product $T_1\cdots T_5$ in the denominator. The distributional (0,4)-form $\bar\delta^{4|4}(T_i\cW^i)$ is defined to be
\be
	\bar\delta^{4,4}\!\left(T_i\cW^i\right)
	:= \delta^{0|4}\!\left(T_i\chi^i\right)\,\prod_{\alpha=0}^3\delbar\left(\frac{1}{T_iW^i_\alpha}\right)\ ,
\label{deltaform}
\ee
where the $\delbar$-operator here is associated with the $T$s (and $\delta^{0|4}(T_i\chi^i) = (T_i\chi^i)^4$ is the standard fermionic $\delta$-function). Since this has homogeneity zero in $T$, the integrand of~\eqref{Rint} is a weightless $(4,4)$-form, so that the integral over $\mathbb{CP}^4$ is well-defined. 

Recalling the standard Cauchy formula
\be
	\delbar \left(\frac{1}{z}\right) = d\bar{z}\,\delta^2(z)\ ,
\label{Cauchy}
\ee
we see that $\bar\delta^{4|4}(T_i\cW^i)$ has support only where
\be
	T_1\cW^1+T_2\cW^2+T_3\cW^3+T_4\cW^4+T_5\cW^5=0
\label{lineardep}
\ee
for each of the supertwistor components -- in other words when the five supertwistors $\cW^1,\ldots,\cW^5$ are linearly dependent. Since $\bar\delta^{4|4}$-function~\eqref{deltaform} has no weight in $\cW$, the result of the integral~\eqref{Rint} will be well-defined on supertwistor space $\mathbb{CP}^{3|4}$ and is manifestly invariant under simultaneous ({\it i.e.} diagonal) superconformal transformations of the five supertwistors. We also note that, just as in~\eqref{Rmomtwist}, equation~\eqref{Rint} treats all five supertwistors on an equal footing, and changes sign under a exchange of any two twistors (because of the corresponding exchange in $T$s).

To prove that~\eqref{Rint} is indeed the usual $R$-invariant, we must perform the integral. This is straightforward -- it is completely fixed by the bosonic $\delta$-functions. Indeed, up to an overall scaling, the $T$s are determined by the bosonic components of~\eqref{lineardep}, as may be seen by contracting with, {\it e.g.}, $\epsilon(\,\cdot\,,2,3,4)$, $\epsilon(1,\,\cdot\,,3,4)$, $\epsilon(1,2,\,\cdot\,,4)$ and $\epsilon(1,2,3,\,\cdot\,)$ respectively. One finds
\be
\begin{aligned}
	\bar\delta^4(T_iW^i) &= \frac{1}{T_5^4\epsilon(1,2,3,4)}
	\bar\delta\!\left(\frac{T_1}{T_5}-\frac{\epsilon(2,3,4,5)}{\epsilon(1,2,3,4)}\right)\,
	\bar\delta\!\left(\frac{T_2}{T_5}-\frac{\epsilon(3,4,5,1)}{\epsilon(1,2,3,4)}\right)\\
	&\hspace{3.5cm}\times\ \bar\delta\!\left(\frac{T_3}{T_5}-\frac{\epsilon(4,5,1,2)}{\epsilon(1,2,3,4)}\right)\,
	\bar\delta\!\left(\frac{T_4}{T_5}-\frac{\epsilon(5,1,2,3)}{\epsilon(1,2,3,4)}\right)
\label{Tfix}
\end{aligned}
\ee
with the factor of $T_5^4\epsilon(1,2,3,4)$ arising as a Jacobian. The ratios $T_j/T_5$ (for $j=1,\ldots,4$) are thus fixed, with the underlying geometric interpretation that linear dependence of five \emph{bosonic} twistors is automatic -- we can always use four twistors (in general position) as a basis of $\mathbb{CP}^3$. The ratios $T_j/T_5$ are simply the components of $W^5$ in the $\{W^1,\ldots,W^4\}$ basis.

The remaining fermionic $\delta^{0|4}$-function keeps track of the fact that linear dependence is not automatic for five \emph{super}twistors, and constrains their anticommuting parts to have the \emph{same} component decompositions as the commuting parts. Substituting the ratios $T_j/T_5$ from~\eqref{Tfix} into  $\delta^{0|4}(T_i\chi^i)$, the overall factors of $T_5$  cancel (as they must for homogeneity) and one recovers
\be
	\int_{\mathbb{CP}^4} \frac{D^4T}{T_1\cdots T_5} \,\bar\delta^{4|4}\!\left(\sum_{i=1}^5 T_i\cW^i\right)
	= \frac{\delta^{0|4}\!\left(\chi^5\,\epsilon(1,2,3,4) + \hbox{cyclic}\right)}
	{\epsilon(1,2,3,4)\epsilon(2,3,4,5)\epsilon(3,4,5,1)\epsilon(4,5,1,2)\epsilon(5,1,2,3)}
\label{Rintreal}
\ee
as promised. 

We have thus written the basic $R$-invariants in a way that makes their dual superconformal invariance manifest -- without reference to any choice of space-time signature. More importantly, we see they have the simple geometric interpretation as the condition that five momentum supertwistors be linearly dependent. The fact that the basic dual superconformal invariants $R_{t;rs}$ of~\eqref{Rdef} can be given such an elegant description as the condition that five supertwistors be linearly dependent is an illustration of the usefulness of the (momentum) twistor representation when dealing with (dual) conformal objects. In~\cite{Hodges:2009hk}, Hodges gave an alternative interpretation of $R_{1,5}$ as the volume of a polytope in momentum supertwistor space. The relation of Hodges' picture to ours will be explored in section~\ref{sec:polytopes}.

\bigskip

For some purposes, it will be useful to represent $R_{1,5}$ as a \emph{contour} integral
\be
	R_{1,5}=\frac{1}{(2\pi\im)^4}\oint_\Gamma \frac{D^4T}{T_1\cdots T_5}\,
	\frac{\delta^{0|4}\!\left(T_i\chi^i\right)}{\prod_{\alpha=0}^3\left(T_iW^i_\alpha\right)}\ ,
\label{Rcdef}
\ee
where the contour is chosen to encircle each of the four simple poles in the product over bosonic twistor components, {\it i.e.}
\be
	\Gamma=\left\{T\in\mathbb{CP}^4\  \hbox{such that} \  
	\left| \sum_{i=1}^5T_iW^i_\alpha\right| = \varepsilon_\alpha \right\}
\label{contour}
\ee
for some positive infinitesimals $\varepsilon_\alpha$.  As is standard (see {\it e.g.}~\cite{GH,T}), we orient $\Gamma\simeq (S^1)^4$ by the condition
\be
	d\phi_0\wedge d\phi_1\wedge d\phi_2\wedge d\phi_3 \geq0\ ,
\ee
where $\phi_\alpha = {\rm arg}\left(T_iW^i_\alpha\right)$. Because of the ordering in $\alpha$, this orientation depends on the orientation of the twistor space. In particular, the orientation of $\Gamma$ is reversed -- and~\eqref{Rcdef} changes sign -- if any two twistors are interchanged, since any basis of the twistor space $\mathbb{CP}^3$ that includes the two twistors in question will also change orientation. $\Gamma$ restricts the support of the $T$s in just the same way as the bosonic $\delta$-functions of~\eqref{deltaform} and the contour integral may be performed using the same change of variables as in~\eqref{Tfix}. Note that the contour version~\eqref{Rcdef} may be derived directly from the Dolbeault form~\eqref{Rint} using Stokes' theorem (see {\it e.g.}~\cite{GH}).

In the rest of the paper,  it will be convenient to write
\be
	R_{1,5} = \frac{1}{(2\pi\im)^4}\oint_\Gamma \frac{D^4T}{T_1\cdots T_5} \ \delta^{4|4}\!\left(\sum_iT_i\cW^i\right)
\label{Rcont}
\ee
even with complex momentum twistors, so as to emphasize the dual superconformal invariance. This notation is used with the understanding that the bosonic $\delta$-functions are to be treated as Cauchy poles with the contour~\eqref{contour}.


\subsection{Real twistor formulations}
\label{sec:realtwistors}

In the case of (2,2) signature space-time, the (momentum) twistors are
completely real and one may be tempted to interpret~\eqref{Rcont} as
an integral of a real $\delta$-function over $\mathbb{RP}^4$. This is
not quite right --  $\mathbb{RP}^4$ is not orientable, so $D^4T/ T_1\cdots T_5$ does not provide a integral density\footnote{The scaling relation $[T]\sim [rT]$ on real projective spaces has
  $r\in\mathbb{R^*}$. This has two connected components,
  $r\in\mathbb{R}^\pm$, and positive and negative scalings must be
  considered separately. Quotienting by the positive scalings yields
  the sphere $S^n$.  The form $D^nT/ T_1\cdots T_n$ is invariant
  under a negative scaling, but the orientation of $S^n$ reverses when
  $n$ is even.}.  However, $D^4T/|T_1\cdots T_4|$ does provide a
density that can be integrated over $\mathbb{RP}^4$ globally.  There is
also an extra modulus sign in the Jacobian of the change of
variables~\eqref{Tfix} in the $\delta$-functions, corresponding to the
familiar \be \delta(ax) = \frac{1}{|a|}\delta(x) \qquad\hbox{instead
  of}\qquad \bar\delta(az) = \frac{1}{a}\bar\delta(z)\ .  \ee
Therefore (even with this density) treating~\eqref{Rcont} as a real integral yields $R_{1,5}$, but with an overall modulus sign in the denominator 
\be
	\int_{\mathbb{RP}^4} \frac{D^4T}{|T_1\cdots T_5|} \ \delta^{4|4}\!\left(\sum_iT_i\cW^i\right)  =
	\frac{\delta^{0|4}\!\left(\chi^5\,\epsilon(1,2,3,4) +  \hbox{cyclic}\right)} 
	{|\epsilon(1,2,3,4)\epsilon(2,3,4,5) \epsilon(3,4,5,1)\epsilon(4,5,1,2) \epsilon(5,1,2,3)|}  
\label{Rcont-split}
\ee
and one needs provide an additional overall sign to recover the correct result.  One such correct formula is
\be
R_{1,5}=\mathrm{sgn} (\epsilon(2,3,4,5) )\int \frac{\mathrm{d}^4 t}{t_1\cdots
  t_4} \ 
\delta^{4|4}\!\left(W_1+\sum_{i=2}^4t_i\cW^i\right) 
\label{Rcont-split-a}
\ee
Of course, the contour integral~\eqref{Rcont} applies in any signature, or in complex momentum space. In split signature, one would simple find that all the poles happen to lie on a copy of $\mathbb{RP}^4\subset\mathbb{CP}^4$.

\bigskip

Similar awkward signs arise in the context of analyzing scattering
amplitudes using Witten's half-Fourier transform to usual twistor
space ({\it i.e.} the space of the fundamental representation of the
usual superconformal group) as found in~\cite{Mason:2009sa,ArkaniHamed:2009si,Korchemsky:2009jv}.  For many purposes
the split signature formulation yields significant insights, but it has
always been encumbered with these sign defects.  These are tied to the
fact that cohomology is sidestepped in this formulation, and so
cohomological signs have to re-appear somewhere in order to get the
right answer.  The signs are incorporated naturally by the use of
contour integrals or Dolbeault $\bar\delta$-functions as above.


\section{Grassmannians and Momentum Twistors}
\label{sec:Grassmannian}

Having interpreted the simplest dual superconformal invariant, various questions naturally arise. Firstly, can one understand how to naturally obtain \emph{sums} of these basic $R$-invariants, as required for example in NMHV tree amplitudes or certain box coefficients? Secondly, can one find a similarly natural expression for the higher-order $R$-invariants that appear at N$^2$MHV level and beyond?

The key to answering these and other questions is to interpret~\eqref{Rcont} as a particular case of a contour integral in the Grassmannian of $k$-planes in $\mathbb{C}^n$, ${\rm G}(k,n)$  (equation~\eqref{Rcont} being the case $k=1$, $n=5$ for which ${\rm G}(1,5) =\mathbb{CP}^4$).  The idea to interpret (objects related to) the S-matrix of $\mathcal{N}=4$ SYM in terms of such a contour integral over a Grassmannian was introduced by Arkani-Hamed {\it et al.}~\cite{ArkaniHamed:2009dn} in the context of usual supertwistors. Our use of the Grassmannian is directly inspired by ref.~\cite{ArkaniHamed:2009dn}, but instead works with momentum twistors.

To motivate the Grassmannian integral, first consider extending~\eqref{Rcont} to the $n$-particle contour integral
\be
	R_{1,n}:=\frac{1}{(2\pi\im)^{n-1}}\oint\frac{D^{n-1}T}{T_1\cdots T_n}\, \delta^{4|4}\!\left({\bf T\cdot W}\right)\ .
\label{ncont}
\ee
where ${\bf T} = (T_1,\ldots, T_n)$ determines a line in $\mathbb{C}^n$, and we have assembled the $n$ supertwistors into the column vector
\be
	{\bf W} = \begin{pmatrix} 
				\cW^1\\
				\cW^2\\
				\vdots\\
				\cW^n
			\end{pmatrix}
\ee
so that ${\bf T\cdot W} = \sum_{i=1}^n T_i\cW^i$.  	
	
The integrand of~\eqref{ncont} is a meromorphic top-degree form on $\mathbb{CP}^{n-1}$, but as there are still only four bosonic $\delta$-functions (really, Cauchy poles associated with a $(S^1)^4$ contour), we must still choose an $n-5$ dimensional contour in order for~\eqref{ncont} to be meaningful. We can recover $R_{1,5}$ by choosing this remaining contour to encircle each of the $n-5$ simple poles $T_6,\ldots,T_n$ in the measure: when $T_6,\ldots, T_n$ vanish, \eqref{ncont} becomes independent of twistors $\cW^6,\ldots,\cW^n$. It should now be clear that (just as in~\cite{ArkaniHamed:2009dn}) we will be able to extract \emph{sums} of basic $R$-invariants by choosing a contour that encircles an appropriate combination of poles from the measure of~\eqref{ncont}, with the sum arising from the usual sum over residues at the various singularities. Moreover, by deforming the contour and using higher-dimensional versions of Cauchy's theorem,
\be
	 \sum \hbox{Res}\  \omega = 0
\ee
for $\omega$ a meromorphic 1-form on $\mathbb{CP}^1$, we can obtain a host of identities relating sums of different $R$-invariants. In the context of usual twistor space (or momentum space), a detailed discussion of the appropriate contours for $n$-particle NMHV tree amplitudes and box coefficients was given by Arkani-Hamed {\it et al.}~\cite{ArkaniHamed:2009dn}. We will see in section~\ref{sec:contours} that the closely related contours are also appropriate in the context of momentum twistors.

The original, twistorial Grassmannian picture of~\cite{ArkaniHamed:2009dn} obtains NMHV amplitudes and box coefficients from a contour integral in G$(3,n)$ rather than our projective space G$(1,n)$. Likewise, the denominator $T_1\ldots T_n$ of~\eqref{ncont} is replaced in~\cite{ArkaniHamed:2009dn} by an $n$-fold product of $2\times2$ determinants. The added simplicity of the momentum twistor approach here comes at the expense of not reproducing the MHV prefactor that appears in the amplitude and is obtained in~\cite{ArkaniHamed:2009dn}. In this respect, the momentum twistor approach is somewhat complementary to the polygonal Wilson loop of~\cite{Drummond:2007aua,Brandhuber:2007yx,Drummond:2008aq,Anastasiou:2009kna} that yields \emph{only} the MHV prefactor.

\bigskip

To consider N$^k$MHV amplitudes, we generalize~\eqref{ncont} to 
\be
	R_{k,n}:=\frac{1}{(2\pi\im)^{k(n-k)}} \oint_{\Gamma\subset {\rm G}(k,n)}\hspace{-0.7cm}d\mu\hspace{0.6cm} 
	\prod_{r=1}^k\ \delta^{4|4}\!\left({\bf T}^r{\bf\cdot W}\right)\ .
\label{G}
\ee
For each $r$, ${\bf T}^r$ is a vector in $\mathbb{C}^n$, and the $k$ such vectors 
\be
	\begin{pmatrix}
		{\bf T}^1\\
		\vdots\\
		{\bf T}^k
	\end{pmatrix}
	=
	\begin{pmatrix}
		T^1_{\ 1}& T^1_{\ 2} & \cdots & T^1_{\ n\!-\!1} &T^1_{\ n}\\
		\vdots &\vdots &  & \vdots & \vdots\\
		T^k_{\ 1}& T^k_{\ 2} & \cdots & T^k_{\ n\!-\!1} & T^k_{\ n}
	\end{pmatrix}
\ee
determine a $k$-plane. The $T^r_{\ i}$ are the $k\times n$ homogeneous coordinates on G$(k,n)$. Note that
\be
	{\rm dim}_{\mathbb C} \,{\rm G}(k,n) = k(n-k)\ .
\label{dimG}
\ee
The measure $d\mu$ is defined as follows. Firstly, there is a natural holomorphic $k(n-k)$-form that is $GL(k)\times GL(n)$ invariant:
\be
	D^{k(n-k)}T \equiv T^{i_{1\,1}\ldots i_{1\,(n-k)}}\cdots T^{i_{k\,1}\ldots i_{k\,(n-k)}}\,
	(d^kT)_{i_{1\,1}\ldots i_{k\,1}}\wedge\ldots\wedge(d^kT)_{i_{1\,(n-k)}\ldots i_{k\,(n-k)}}
\label{canonGrass}
\ee
where
\be
	T^{i_1\ldots i_{n-k}} \equiv \frac{1}{n!k!}
	\epsilon^{i_1\ldots i_n} \epsilon_{r_1\ldots r_k} T^{r_1}_{\ i_{(n-k+1)}}\cdots T^{r_k}_{\ i_n}
	\quad\hbox{and}\quad
	(d^kT)_{i_{1}\ldots i_{k}} \equiv \frac{1}{k!}
	\epsilon_{r_1\ldots r_k} dT^{r_1}_{\ i_1}\wedge\ldots\wedge dT^{r_k}_{\ i_k}\ .
\ee
$D^{k(n-k)}T$ is thus a natural generalization of the weighted top holomorphic form
\be
	D^{n-1}T \equiv \frac{1}{n!}\epsilon^{i_1\ldots i_n} T_{i_1}dT_{i_2}\wedge\ldots\wedge dT_{i_n}
\ee
on $\mathbb{CP}^{n-1}$, and is invariant under \emph{local} $GL(k)$ transformations ({\it i.e.} ones that vary over G$(k,n)$). One can fix this $GL(k)$ freedom by reducing a $k\times k$ block of $T$ to the identity matrix, {\it e.g.} on a patch of G$(k,n)$ where the determinant of the first $k$ columns of $T$ is non-vanishing, we can set
\be
	\begin{pmatrix}
		T^1_{\ 1}& T^1_{\ 2} & \cdots & T^1_{\ k} & T^1_{\ k\!+\!1} & \cdots &T^1_{\ n}\\
		\vdots && &\vdots & \vdots & & \vdots\\
		T^k_{\ 1}& T^k_{\ 2} & \cdots & T^k_{\ k} & T^k_{\ k\!+\!1} & \cdots & T^k_{\ n}
	\end{pmatrix}
	\buildrel{GL(k)}\over\longrightarrow
	\begin{pmatrix}
		1 & \cdots & 0 & t_{1\,k\!+\!1} & \cdots & t_{1\, n}\\
		\vdots &\ddots & \vdots & \vdots & & \vdots\\
		0 & \dots & 1 & t_{k\, k\!+\!1} & \cdots & t_{k\, n}
	\end{pmatrix}\ .
\ee		
whereupon~\eqref{canonGrass} reduces to the standard measure $d^{k(n-k)}t$ on our $\mathbb{C}^{k(n-k)}$ coordinate patch.

Just as in projective space, $DT$ has weight $kn$ in the homogeneous coordinates. There is no canonical unweighted measure on G$(k,n)$, and any choice of measure will break the $GL(n)$ symmetry. However, the $\delta$-functions in~\eqref{G} couple $GL(n)$ transformations of the Grassmannian to the $n$ external twistors. From the point of view of colour-ordered planar Yang-Mills amplitudes, the largest symmetry group one expects is the dihedral group on $n$ elements, comprising cyclic permutations and reflections. Now, one \emph{can} make a choice of measure that breaks $GL(n)$ to this dihedral group: following~\cite{ArkaniHamed:2009dn} we define
\be
	d\mu \equiv \frac{D^{k(n-k)}T}{(12\cdots k)(23\cdots k\!+\!1)\cdots (n1\cdots k\!-\!1)}\ ,
\label{measure}
\ee
where $(12\cdots k)$ denotes the $k\times k$ determinant\footnote{In~\cite{ArkaniHamed:2009dn}, the measure associated with an N$^k$MHV amplitude instead involves determinants of $(k+2)\times (k+2)$ matrices.}  formed from the first $k$ columns of  $T$, $(23\cdots k\!+\!1)$ is the $k\times k$ determinant formed from columns 2 to $k+1$, and so forth\footnote{The minors can be viewed as the homogeneous coordinates of the Pl\"ucker embedding G$(k,n)\hookrightarrow \mathbb{CP}^{\frac{n!}{k!(n-k)!}-1}$.}. The $n$-fold product of minors has homogeneity $kn$, so $d\mu$ is a top meromorphic form of homogeneity zero, providing a measure on G$(k,n)$.  Combining this with the superconformal $\delta$-functions shows that the integrand in $R_{k,n}$ is manifestly (dual) superconformally invariant, and is also manifestly symmetric under dihedral permutations of the external particles. Of course, when $k=1$ we recover $R_{1,n}$ as in~\eqref{ncont}.

\bigskip

The $4k$ bosonic $\delta$-functions imply an $(S^1)^{4k}$ contour as before. When $k=n-4$ ($\overline{\rm MHV}$), this suffices to fix the integral completely; complementary to the fact that when $k=0$ (MHV), the ratio
\be
	R_{0,n} = 1
\ee
by definition. In general, though, we must still specify a contour of dimension $k(n-4-k)$ to fix $R_{k,n}$. The bosonic $\delta$-functions restrict the support of~\eqref{G} to the subvariety of G$(k,n)$ where\footnote{A reminder: $\alpha$ indexes the four bosonic components of a twistor.}
\be
	{\bf T}^r\cdot {\bf W}_\alpha = 0\ ,
\label{restrict}
\ee
so that geometrically, each of the vectors ${\bf T}^r$ must be orthogonal to four fixed vectors ${\bf W}_\alpha$ whose direction depends only on the external twistors. Rather than $k$-planes in $\mathbb{C}^n$, the orthogonality conditions~\eqref{restrict} imply that the $k$-planes spanned by ${\bf T}^r$ are constrained to lie in a $\mathbb{C}^{n-4}\subset\mathbb{C}^n$ -- in other words, the bosonic $\delta$-functions reduce the support of $R_{k,n}$ to a G$(k,n-4)$ linearly embedded in G$(k,n)$, of dimension $k(n-4-k)$. The fact that after accounting for momentum conservation the `true' Grassmannian is\footnote{Note that $k_{\rm here} = k_{\rm there}-2$.} G$(k,n-4)$ was also noted in ref.~\cite{ArkaniHamed:2009dn}, but in \emph{momentum} twistor space one considers an embedding 
\be
	{\rm G}(k,n-4)\hookrightarrow{\rm G}(k,n)
\ee
whereas in twistor space one has
\be
	{\rm G}(k,n-4)\hookrightarrow{\rm G}(k+2,n)
\ee 
instead.

A detailed examination of various choices of contour follows in section~\ref{sec:contours}. The first case that requires the full machinery of the Grassmannian (rather than just projective space) is $k=2$ or N$^2$MHV, where we must still choose a $2(n-6)$ dimensional contour. Again, we find that the required contours are closely related to those that were relevant in~\cite{ArkaniHamed:2009dn}. As in projective space, higher-dimensional generalizations of Cauchy's theorem allow us to obtain sums of terms of the appropriate R-charge by choosing contours that enclose several poles in $d\mu$. Contour deformation arguments will then provide us with myriad identities relating different forms of the sums of coefficients. All of this may be done in a way that keeps dual superconformal symmetry manifest throughout.


\section{Examples of Contours}
\label{sec:contours}

In this section, we will illustrate the specification of contours necessary to extract various NMHV \& N$^2$MHV tree amplitudes and box coeefficients (all divided by the MHV tree amplitude) from the Grassmannian master formula
\be
	R_{k,n} = \frac{1}{(2\pi\im)^{k(n-k)}}\oint d\mu\ \prod_{r=1}^k\delta^{4|4}\!\left({\bf T}^r{\bf \cdot W}\right)
\label{G2}
\ee
on momentum twistor space. Our aim here is not to give a comprehensive list of all possible contour choices, nor even to 
demonstrate all the remarkable properties such multi-dimensional contour integrals possess (although a modest example - the equivalence of different BCFW decompositions of the 6 particle NMHV tree amplitude will be seen in section~\ref{sec:NMHVtrees}). Rather, we intend to establish that the contour specifications used in the original, superconformal formula of~\cite{ArkaniHamed:2009dn} closely correspond to the appropriate contours in this \emph{dual} superconformal context. 


\subsection{NMHV 3-mass \& 2-mass-hard box coefficients}
\label{sec:NMHVbox}

The simplest case is the 3-mass box function, whose coefficient\footnote{Since our formulation is manifestly dual superconformal invariant, it is clear that we should expect to compute coefficients of box \emph{functions} rather than box \emph{integrals}. Thus this and later figures are somewhat schematic -- the {\it rhs} can only be intepreted as the usual product of tree amplitudes (summed over the internal supermultiplet) once one accounts for this Jacobian (and, indeed, the MHV tree amplitude).}
\begin{figure}[h]
\begin{center}
	\includegraphics[height=50mm]{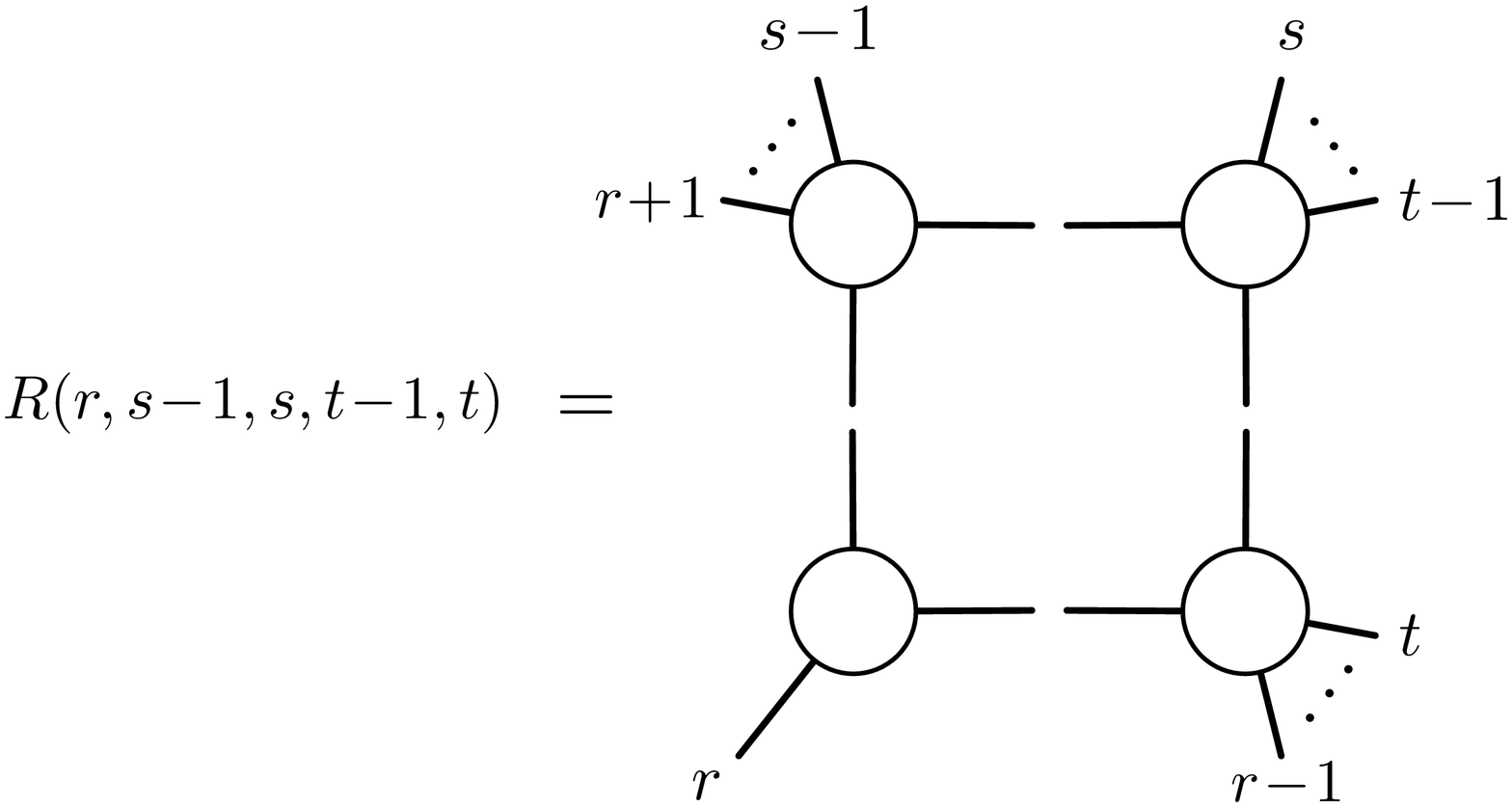}
\end{center}
\end{figure}

\noindent is precisely the basic dual superconformal invariant
obtained from
\be
	R_{1,n}=\oint d\mu \,\delta^{4|4}({\bf T\cdot W})\ .
\ee
To specify the contour, recall that taking the residue at $T_i=0$ ensures that $\cW^i$ drops out of the remaining integral. Hence the appropriate $(n-5)$ dimensional contour (in addition to the 4 $\delta$-functions) should simply be chosen to separately encircle each of the hyperplanes $T_i=0$ for all $i$ except $i\in\{r,s\!-\!1,s,t\!-\!1,t\}$.

The 2-mass-hard box contributions are also straightforward to obtain. They may be thought of as degenerations of the 3-mass box when either 
\be
	s=r+2\qquad\hbox{or}\qquad t=r-1\ ,
\label{3to2mhdeg}
\ee
so that one of the two MHV amplitudes adjacent to the three-particle $\overline{\rm MHV}$ subamplitude itself involves only three particles. Combining them, the coefficient of a single 2-mass-hard box integral is
\begin{figure}[h]
\begin{center}
	\includegraphics[height=55mm]{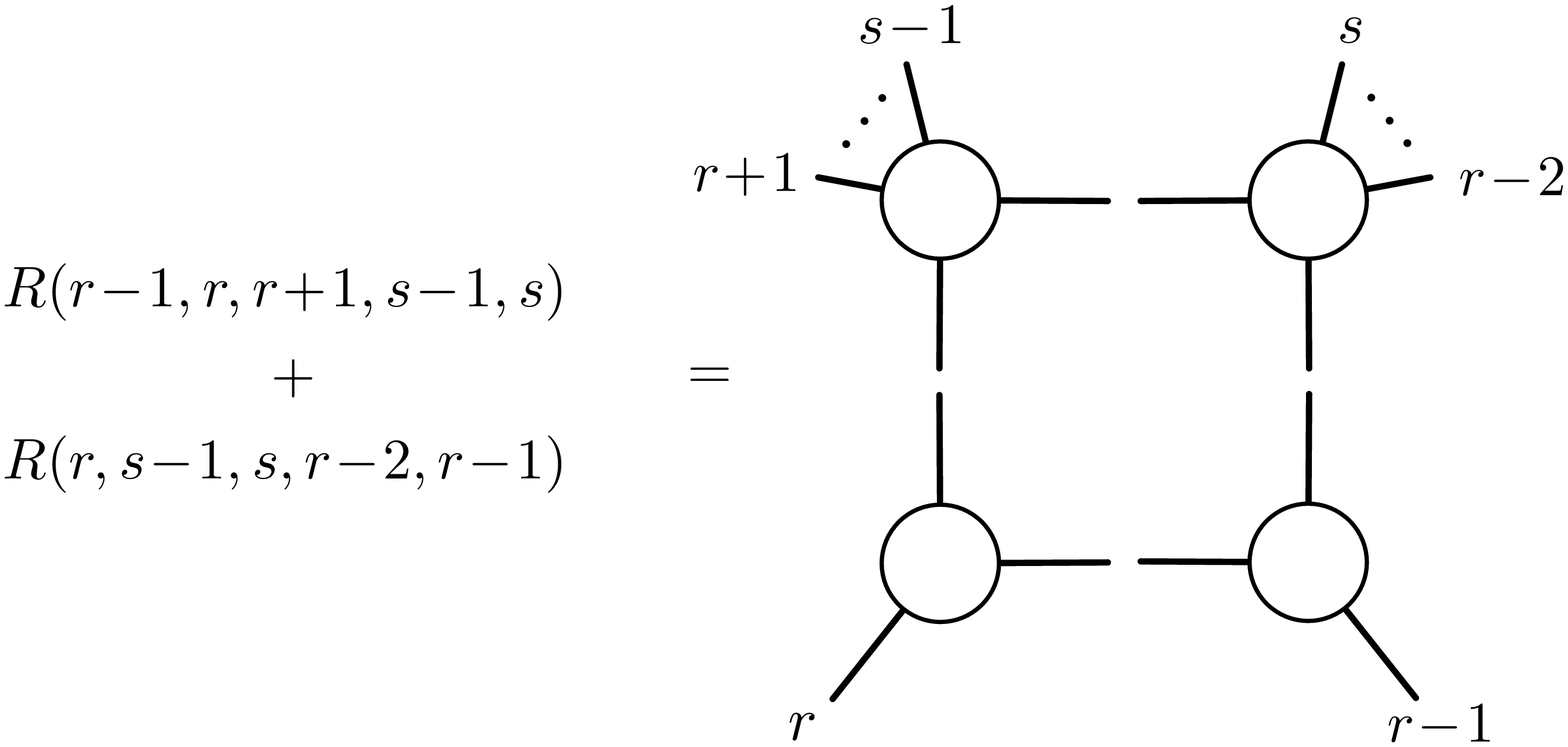}
\end{center}
\end{figure}

\noindent This sum is obtained by first restricting to the $\mathbb{CP}^1\subset\mathbb{CP}^{n-1}$ defined by
\be
	{\bf T\cdot W}_\alpha = 0\ ,\qquad T_i = 0 \quad\hbox{for}\quad i\notin \{r\!-\!1,r,r\!+\!1,r\!+\!2,s\!-\!1,s\}\ ,
\label{2mhCP1}
\ee
and then choosing the remaining $S^1$ contour factor to encircle both the poles $T_{r-1}=0$ and $T_{r+2}=0$.

In~\cite{ArkaniHamed:2009dn}, the 3-mass box coefficient was identified with the contour that computes the residue when all except the $\{r\!-\!1,s\!-\!2,s\!-\!1,t\!-\!2,t\!-\!1\}^{\rm th}$ minors vanish. Likewise, the 2mh box coefficient there came from a contour encircling each of the poles associated to all the minors except $\{r\!-\!1,r,s\!-\!2,s\!-\!1\}$ and either $r\!-\!2)$ or $r\!+\!1$. These contour specifications differ from ours (for the same arrangement of legs on the associated box function) only by a cyclic shift $i\to i\!-\!1$.


\subsection{NMHV tree amplitudes}
\label{sec:NMHVtrees}

The 6-particle NMHV tree amplitude is given by
\be
\begin{aligned}
	\frac{A^{\rm NMHV}_{6,0}}{A^{\rm MHV}_{6,0}} &= R_{6;24}+R_{6;25}+R_{6;35}\\
	&=R(6,1,2,3,4)+R(6,1,2,4,5)+R(6,2,3,4,5)\ .
\end{aligned}
\ee
This sum of dual conformal invariants may be obtained from~\eqref{Rcont} using the contour that encircles the residues at $T_1 =0$, $T_3=0$ and $T_5=0$ on the $\mathbb{CP}^1\subset\mathbb{CP}^5$ given by ${\bf T\cdot W}_\alpha=0$:
\begin{figure}[h]
\begin{center}
	\includegraphics[height=45mm]{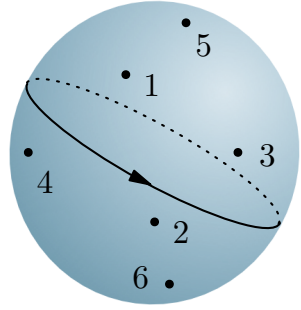}
\end{center}
\caption{{\it The Riemann sphere given by $\{ {\bf T\cdot W}_\alpha=0\}\subset\mathbb{CP}^5$. The six marked points are the intersections of this $\mathbb{CP}^1$ with the hyperplanes $T_i=0$. The homology class of the displayed contour is invariant (up to a reversal in orientation) under cyclic permutations of the external states.}}
\label{fig:6ptcontour}
\end{figure}

\noindent Moreover, the identity
\be
	R_{6;24}+R_{6;25}+R_{6;35} +R_{1;46}+R_{1;36}+R_{1;35} =0
\label{6NMHVcyclic}
\ee
that guarantees cyclic symmetry of the amplitude is manifest with this contour choice, arising as a simple application of Cauchy's theorem.

Singularities of the six particle amplitude arise when the contour becomes pinched -- that is, when a pole in the `upper' and a pole in the `lower' hemisphere of $\{{\bf T\cdot W}_\alpha=0\}$ collide so that ({\it e.g.}) the intersection 
$$
	\{T_1=0\}\cap\{T_4=0\}\cap \{{\bf T\cdot W}_\alpha=0\}
$$
is non-empty. This implies
\be
	T_2W^2+T_3W^3+T_5W^5+T_6W^6 = 0
\ee
so the twistors $\{2,3,5,6\}$  must be coplanar. It then follows that
\be
	x_{36}^2 \propto \epsilon(2,3,5,6) =0
\ee
corresponding to the physical singularity $(p_3+p_4+p_5)^2\to0$ in momentum space. On the other hand, if twistors $\{2,3,4,6\}$ are coplanar so that
$$
	\{T_1=0\}\cap\{T_5=0\}\cap\{{\bf T\cdot W}_\alpha=0\}\neq\emptyset\ ,
$$
the contour integral is clearly unaffected. This coplanarity condition corresponds to the momentum space singularity $\la 6|x_{63}x_{34}|3\ra\to0$, which is spurious.

\bigskip

The six particle tree is  the simplest non-trivial example of an NMHV tree amplitude that requires a choice of contour. A  contour (on the Grassmannian in ordinary twistor space) that is appropriate for general NMHV tree amplitudes was identified in~\cite{ArkaniHamed:2009dn}. In their notation, the $(S^1)^m$ contour that computes the residue when each of the first $m$ minors vanishes is denoted $\{1\}\{2\}\{3\}\cdots\{m\}$, while the sum $\{1\}+\{2\}+\cdots+\{m\}$ denotes a single $S^1$ contour factor that encircles the vanishing locus of all of the first $m$ minors. The NMHV tree amplitudes were then shown to be associated to the contour defined by
\be
	\Gamma^{\rm NMHV}_{\rm tree}\equiv
	\underbrace{\ \mathcal{E}\star\mathcal{O}\star\mathcal{E}\star\cdots\ }_{(n-5)\ \rm{factors}}
\label{NMHVtreecont}
\ee
where
\be
	\mathcal{E}\equiv \sum_{k\ \rm{even}} \{k\}\qquad\hbox{and}\qquad
	\mathcal{O}\equiv \sum_{k\ \rm{odd}} \{k\}
\ee
and the star product is defined as
\be
	\{i_1\}\star\{i_2\} \equiv
	\begin{cases}
		\{i_1\}\{i_2\} & \hbox{if\ } i_1< i_2\\
		\hspace{0.65cm} 0 & \hbox{otherwise}.
	\end{cases}
\label{stardef}
\ee 
A small amount of experimentation is enough to convince oneself that the same contour prescription correctly reproduces $A^{\rm NMHV}_{n,0} / A^{\rm MHV}_{n,0}$ from the momentum twistor Grassmannian. As a second example, when $n=8$, $\Gamma^{\rm NMHV}_{\rm tree}$ becomes
\be
\begin{aligned}
	\Gamma^{\rm NMHV}_{8} &= \{2\}\{3\}\{4\}+\{2\}\{3\}\{6\}+\{2\}\{3\}\{8\} + \{2\}\{5\}\{6\}+\{2\}\{5\}\{8\}+\{2\}\{7\}\{8\}\\
	&\qquad +\{4\}\{5\}\{6\}+\{4\}\{5\}\{8\}+\{4\}\{7\}\{8\}+\{6\}\{7\}\{8\}
\end{aligned}
\ee
and  yields
\be
\begin{aligned}
	&R(1,5,6,7,8)+R(1,4,5,7,8)+R(1,4,5,6,7)+R(1,3,4,7,8)+R(1,3,4,6,7)\\
	&\qquad +R(1,3,4,5,6)+R(1,2,3,7,8)+R(1,2,3,6,7)+R(1,2,3,5,6)+R(1,2,3,4,5)\\
	&\qquad\qquad= \sum_{3\leq a<b-1\leq7} R_{1;ab}\ .
\end{aligned}
\ee
In general,  \eqref{NMHVtreecont} contains $\frac{1}{2}(n-3)(n-4)$ summands, each of which sets $n-5$ homogeneous coordinates to zero. Under a cyclic permutation of the external legs, $\mathcal{E}\leftrightarrow\mathcal{O}$ and~\cite{ArkaniHamed:2009dn} showed that (up to a possible reversal in orientation) the homology class of this contour (and hence the sum of residues) was unchanged. (The cyclic invariance of the tree amplitudes means that the shift $i\to i\!-\!1$ in contour prescriptions noticed above for the box coefficients is irrelevant here.)


\subsection{Second-order invariants for N$^2$MHV amplitudes}
\label{sec:NNMHVtrees}

N$^2$MHV amplitudes are the first case where the full Grassmannian formula
\be
	\frac{1}{(2\pi\im)^{2(n-2)}}\oint d\mu\ \prod_{r=1}^2\ \delta^{4|4}\!\left({\bf T}^r\cdot{\bf W}\right)
\ee
is needed (rather than just projective space).  Not coincidentally, in momentum space they also require the introduction of new, `second-order' objects $R_{n;ab;cd}$, defined as
\be
	R_{n;ab;cd} \equiv \frac{\la c\,c\!-\!1\ra\la d\,d\!-\!1\ra\ 
	\delta^{0|4}\!\left(\la \xi|x_{bc}x_{cd}|\theta_{db}\ra + \la \xi|x_{bd}x_{dc}|\theta_{cb}\ra\right)}
	{x^2_{cd}\la \xi|x_{bc}x_{cd}|d\ra\la \xi|x_{bc}x_{cd}|d\!-\!1\ra\la \xi|x_{bd}x_{dc}|c\ra\la \xi|x_{bd}x_{dc}|c\!-\!1\ra}\ .
\label{2Rdef}
\ee
with 
\be
	\la\xi|\equiv\la n|x_{na}x_{ab}\ .
\label{xidef}
\ee
These second-order $R$s are dual conformally invariant, but only become dual \emph{super}conformally invariant on the support of an appropriate first-order $R$. Accordingly, the N$^2$MHV tree amplitude may be writted as~\cite{Drummond:2008cr} 
\be
	\frac{A^{{\rm N}^2{\rm MHV}}_{n,0}}{A^{\rm MHV}_{n,0}} 
	= \sum_{2\leq a,b<n} R_{n;ab} \left(\sum_{a< c,d\leq b} R^{ab}_{n;ab;cd}\ 
	+ \sum_{b\leq c,d<n} R^{ab}_{n;cd}\right)
\label{NNMHVtree}
\ee
where the sums in~\eqref{NNMHVtree} are taken over values of $(a,b)$ and $(c,d)$ in the allowed ranges that also satisfy $a<b\!-\!1$, $c<d\!-\!1$. The superscripts indicate that we should replace
\be
	\la d | \to \la n|x_{na}x_{ab}\qquad\hbox{or}\qquad
	\la c\!-\!1| \to \la n|x_{na}x_{ab}\ ,
\ee
in the boundary cases of $d=b$ or $c=b$ of the first and second sums, respectively. All told, in addition to the basic $R_{n;ab}$ invariants, N$^2$MHV tree amplitudes involve three new objects:
$$
	R_{n;ab;cd}\,,\qquad R^{ba}_{n;ab;ad}\qquad \hbox{and}\qquad R^{ab}_{n;bd}\ ,
$$
at least when written in the form of equation~\eqref{NNMHVtree}. Translating these to momentum twistor space, we find\footnote{See appendix for a derivation.}
\be
\begin{aligned}
	&R_{n;ab;cd} &=&\ R(\cU,c\!-\!1,c,d\!-\!1,d)\,,\\
	&R^{ab}_{n;ab;cb} &=&\ R(\cU,c\!-\!1,c,b\!-\!1,b)\ \times\ 
	\frac{\epsilon(\cU,c\!-\!1,c,b)\,\epsilon(b\!-\!1,a\!-\!1,a,n)}
	{\epsilon(\cU,c\!-\!1,c,\left[b)\phantom{\int}\hspace{-0.2cm}\epsilon(b\!-\!1\right]\!,a\!-\!1,a,n)}\\
	&R^{ab}_{n;bd}  &=&\ R(n,b\!-\!1,b,d\!-\!1,d)\ \times\ 
	\frac{\epsilon(n,d\!-\!1,d,b\!-\!1)\,\epsilon(b,a\!-\!1,a,n)}
	{\epsilon(n,d\!-\!1,d,\left[b\!-\!1)\phantom{\int}\hspace{-0.2cm}\epsilon(b\right]\!,a\!-\!1,a,n)}
\end{aligned}
\label{2Rtrans}
\ee
where the supertwistor $\cU$ is defined by
\be
	\cU\equiv \epsilon(n,b\!-\!1,b,[a\!-\!1),a]
\label{superUdef}
\ee
and the square brackets in~\eqref{2Rtrans} \&~\eqref{superUdef} denote antisymmetrization in the two twistors.  Of course, either in the momentum representation~\eqref{2Rdef} or the momentum twistor representation~\eqref{2Rtrans}, many identities between second-order $R$s may be inferred from different expressions for the tree amplitude -- indeed, some such identities were used in~\cite{Drummond:2009ge} to present the N$^2$MHV tree amplitude in a slightly different form. However, these identities are even more algebraically intricate than they were at NMHV level! Once again, such identities naturally arise in the Grassmannian formula via the global residue theorem (although we do not investigate this here -- see~\cite{ArkaniHamed:2009dn} for further details).

It is testament to the power of the Grassmannian generating function that it can generate all of these apparently different objects -- each of the objects in~\eqref{2Rtrans}, multiplied by their appropriate first-order $R$, have a common, simple origin in the Grassmannian. The following example is indicative of the general structure. Consider the case of eight particles where G$(k,n)={\rm G}(2,8)$ and has complex dimension 12. Eight of the integrals are fixed by the $\delta$-functions, so we need to specify a four-dimensional contour. Suppose we compute the residue where the third, fourth, sixth and eighth cyclic minors vanish. Working on the $\mathbb{C}^{12}$ coordinate patch 
\be
	\begin{pmatrix}
		{\bf T}^1\\{\bf T}^2
	\end{pmatrix} 
	=
	\begin{pmatrix} 
		1 & t_{12} & t_{13} & t_{14} & 0 & t_{16} & t_{17} & t_{18} \\
		0 & t_{22} & t_{23} & t_{24} & 1 & t_{26} & t_{27} & t_{28}
	\end{pmatrix}\ ,
\label{4mbccoords}
\ee
this residue is located at the intersection
\be
	\{t_{13}t_{24}=0\}\cap\{t_{14}=0\}\cap\{t_{16}t_{27}-t_{17}t_{26}\}=0\cap\{t_{28}=0\}\ .
\ee
The second and fourth conditions simply set two of the $t$s to zero. The first condition is quadratic; since the contour was specified by the vanishing of a polynomial in the $t$s, it encircles both solutions $t_{13}=0$ and $t_{24}=0$ and the residues at these must be summed. The final condition can then be solved for either $t_{27}$ or $t_{17}$, respectively\footnote{For generic external momenta, the residues vanish both $t_{16}t_{27}=0=t_{17}t_{26}$, as the $\delta^{4|4}$-functions then depend on only four twistors that are not coplanar by assumption.}.  Hence, computing the residue at the vanishing of these minors reduces the Grassmannian to
\be
\begin{aligned}
	&\frac{1}{(2\pi\im)^8}\oint \frac{d^8t}{t_{12}t_{16}t_{17}t_{18}\,t_{22}t_{23}t_{24}t_{26}}\ 
	\delta^{4|4}\!\left({\bf A}^1\cdot{\bf W}\right)\, \delta^{4|4}\!\left({\bf A}^2\cdot {\bf W}\right)\\
	&+\ \frac{1}{(2\pi\im)^8}\oint\frac{d^8t}{t_{13}t_{16}t_{18}(t_{12}t_{23}-t_{13}t_{22})t_{22}t_{26}t_{27}}\ 
	\delta^{4|4}\!\left({\bf B}^1\cdot{\bf W}\right)\, \delta^{4|4}\!\left({\bf B}^2\cdot{\bf W}\right)\ ,
\end{aligned}
\ee
where we have defined the matrices
\be
\begin{aligned}
	\begin{pmatrix}
		{\bf A}^1\\{\bf A}^2
	\end{pmatrix}
	&=
	\begin{pmatrix}
		1 & t_{12} & 0 & 0 & 0 & t_{16} & t_{17} & t_{18} \\
		0 & t_{22} & t_{23} & t_{24} & 1 & t_{26} & t_{26}\frac{t_{17}}{t_{16}} & 0
	\end{pmatrix}
	\\
	\begin{pmatrix}
		{\bf B}^1\\{\bf B}^2
	\end{pmatrix}
	&=
	\begin{pmatrix}
		1 & t_{12} & t_{13} & 0 & 0 & t_{16} & t_{16}\frac{t_{27}}{t_{26}} & t_{18} \\
		0 & t_{22} & t_{23} & 0 & 1 & t_{26} & t_{27} & 0
	\end{pmatrix}\ .
\end{aligned}
\ee
If we introduce 
\be
	\cU=\epsilon(8,1,2,\left[6),7\right]\qquad\hbox{and}\qquad \cV=\epsilon(5,2,3,\left[6),7\right]
\label{UVdef}
\ee
the remaining integrals are fixed by the $\delta$-functions, giving
$$
	R(8,1,2,6,7)R(\cU,2,3,4,5) + R(5,6,7,2,3)R(\cV,8,1,2,3)
	\times\frac{\epsilon(\cV,8,1,3)\epsilon(2,6,7,5)}
	{\epsilon(\cV,8,1,\left[3)\epsilon(2\right],6,7,5)}
$$
which we identify as  $R_{8;27}R_{8;27;35} + R_{5;73}R_{5;73;13}^{73}$.

Similarly, computing the residue where the first, second, fifth and sixth minors vanish, {\it i.e.} where
\be
	\{t_{22}=0\}\cap\{t_{12}t_{23}=0\}\cap\{t_{16}=0\}\cap\{t_{17}t_{26}=0\}
\ee
one finds non-vanishing contributions only from the two cases
\be
	t_{22}=t_{23}=t_{16}=t_{17}=0\qquad\hbox{and}\qquad
	t_{22}=t_{12}=t_{16}=t_{26}=0\ .
\ee
The remaining integrals are again fixed by the $\delta$-functions and one obtains the sum of contributions
\be
	R(8,1,2,3,4)R(8,4,5,6,7) = R_{8;24}R_{8;57}
\ee
and
\be
	R(1,3,4,7,8)R(\cU,4,5,7,8)\times 
	\frac{\epsilon(\cU,4,5,8)\epsilon(7,3,4,1)}{\epsilon(\cU,4,5,\left[8)\epsilon(7\right],3,4,1)}
	= R_{1;48}R_{1;48;58}^{48}\ ,
\ee
coming from the first and second solutions, respectively.

These examples show how the Grassmannian generating function unifies the various different types of contribution to N$^2$MHV amplitudes. In particular, if we specialize to pure external gluons with helicity configuration $(1^+,2^-,3^+,4^-,5^+,6^-,7^+,8^-)$, it is readily verified that $R_{8;24}R_{8;35}$ gives (after multiplication by the MHV tree factor) the contribution
$$
	\frac{[13]^4[57]^4\la48\ra^4}{[12][23][56][67]\la 4|2+3|1]\la 8|1+2|3]s_{123}\la 4|5+6|7]\la8|6+7|5]s_{567}}\ ,
$$
where $s_{123} = (p_1+p_2+p_3)^2$ and similarly for $s_{567}$; this is the contribution that~\cite{Britto:2004ap} denoted by $U$. Similarly, the apparently more complicated expression $R_{5;73}R^{73}_{5;73;13}$ reduces to nothing more than (minus) the conjugate term $P(g^4 U)$, where $g:i\to i\!+\!1$ is the cyclic shift and $P$ is the (Minkowski signature) parity operator, exchanging primed and unprimed spinors. Precisely these contributions to the 8-particle alternating helicity amplitude were computed in~\cite{ArkaniHamed:2009dn} from a contour that localises on the vanishing of the first, fourth, fifth and eighth cyclic minors of the twistor Grassmannian formula. As at NMVH, we see there is a cyclic shift
$$
	i_{\rm here} \to {i\!-\!1}_{\rm there}
$$
in the contour prescriptions, for the same labelling of external states. A contour that generates the complete 8-particle N$^2$MHV tree amplitude for this helicity configuration was given in~\cite{ArkaniHamed:2009dn}.


\subsection{The four-mass box coefficient}
\label{sec:NNMHVbox}

A particularly important dual superconformal invariant, that first
becomes relevant at N$^2$MHV, is the coefficient of the 4-mass box
function. This was computed in~\cite{Drummond:2008bq,Hall:2009xg} to
be 
\begin{figure}[h]
\begin{center}
	\includegraphics[height=54mm]{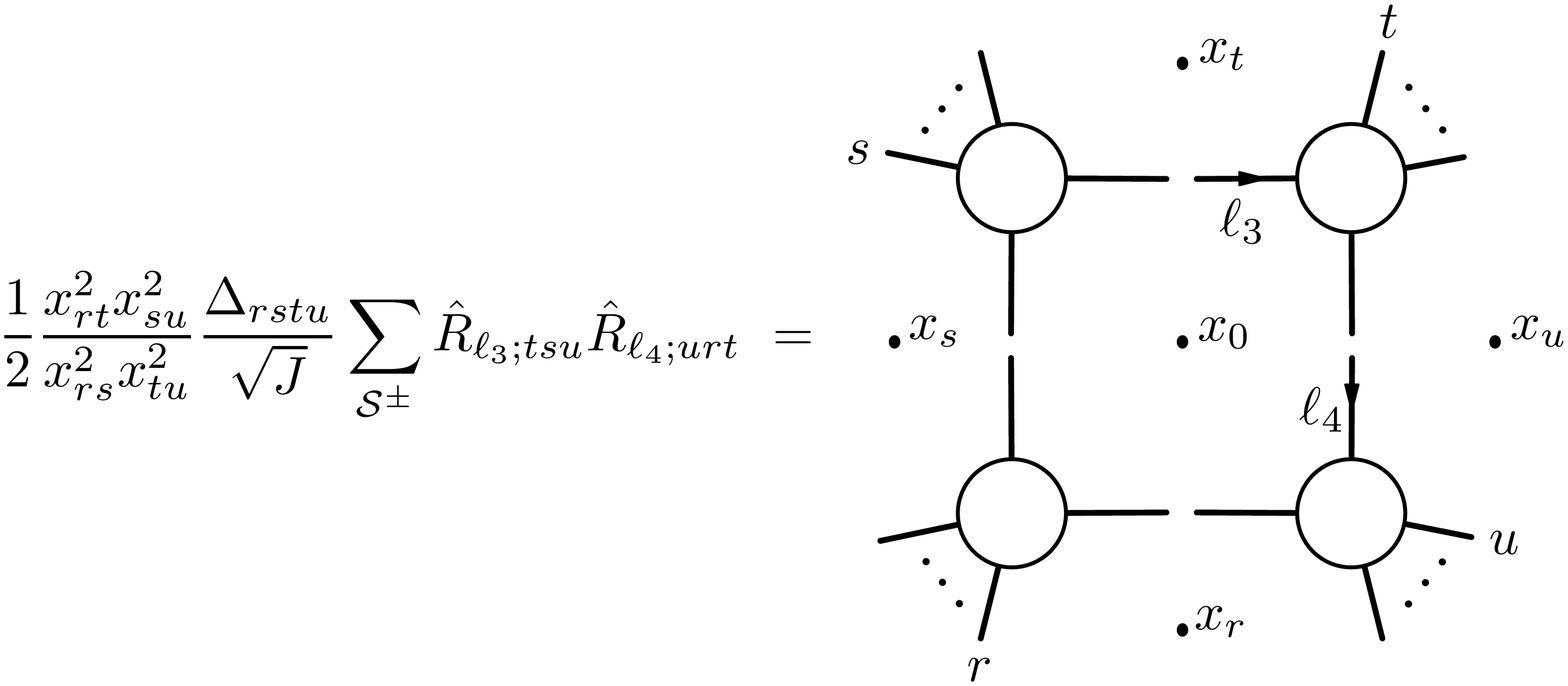}
\end{center}
\caption{\emph{The four-mass box coefficient.}}
\label{4mbox}
\end{figure}

\noindent where (as indicated) $\ell_3$ and $\ell_4$ are the momenta flowing along the (cut) propagators between legs $t\!-\!1$ and $t$, and legs $u\!-\!1$ and $u$, respectively. The sum is taken over the two solutions of the quadruple cut equations
\be
	x_{0r}^2= 0 \ ,\quad x_{0s}^2 = 0\ , \quad x^2_{0t}=0\ , \quad x^2_{0u}=0\ .
\label{quadcut}
\ee
The dual superconformal invariants $\hat R_{\ell_3;tsu}\hat R_{\ell_4;urt}$ depend on the cut loop momenta and were defined in~\cite{Drummond:2008bq} by
\be
	\hat R_{\ell_3;tsu} \equiv \frac{\la s\!-\!1\,s\ra\la u\!-\!1\, u\ra\ \delta^{0|4}(\hat\Xi_{\ell_3;tsu})}
	{x_{su}^2\la\ell_3|x_{tu}x_{us}|s\!-\!1\ra\la\ell_3|x_{tu}x_{us}|s\ra
	\la\ell_3|x_{tu}x_{us}|u\!-\!1\ra\la\ell_3|x_{ts}x_{su}|u\ra}\ ,
\label{loopRdef}
\ee
where
\be
	\hat\Xi_{\ell_3;tsu} \equiv x^2_{su}\la\ell_3|\theta_t\ra + \la\ell_3|x_{ts}x_{su}|\theta_u\ra 
	+ \la\ell_3|x_{tu}x_{us}|\theta_s\ra
\ee
and similarly for $\hat R_{\ell_4;urt}$. 

As is well-known, the quadruple cut of the 4-mass box fixes the loop
momenta in terms of a quadratic expression in the external momenta,
so~\eqref{loopRdef} is not a rational function of the external spinors
(although the overall box coefficient itself is). The presence of the
radicals in $\hat{R}_{\ell_3;tsu}$ and $\hat{R}_{\ell_4;urt}$ is a
tell-tale sign of the 4-mass box coefficient, providing a clear signal
that one is studying a leading singularity of a \emph{loop}
expression, rather than merely a combination of terms in the BCFW
decomposition of a tree. It is thus particularly important to show
that these quadratic expressions can be obtained from the dual
superconformal Grassmannian in momentum twistor space. We now examine
this in some detail. Although we do not derive one of the standard
formulae for the required box coefficient from our Grassmannian
formula, there are many different representations of such terms
arising from different choices of ordering in solving the 
equations that arise, and
we believe our formula to be just such a new representation.

\bigskip

In the simplest case of eight particles (with $r = 2$ in the above
diagram), \cite{ArkaniHamed:2009dn} identified this box coefficient as
the residue of a contour encirling the subvariety where the even cyclic minors vanish. Once again, we will
show that an analogous contour is appropriate also in momentum
twistors. Note that because~\cite{ArkaniHamed:2009dn} were dealing
with $4\times 4$ determinants, the residues from the Grassmannian
formula could only be computed numerically; here the minors are
$2\times 2$ determinants, making it feasible to do the calculation
analytically, thus clarifying much of the geometric structure.

Working on the same coordinate patch~\eqref{4mbccoords} as in section~\ref{sec:NNMHVtrees}, choose the contour that localises the integrand on the subvariety where the \emph{odd} cyclic minors
$$
	t_{22}\, ,\quad (t_{13}t_{24}-t_{14}t_{23})\, ,\quad t_{16}\quad\hbox{and}\quad
	 (t_{17}t_{28}-t_{18}t_{27})
$$
vanish. The Grassmannian integral reduces to
\be
	\frac{1}{(2\pi\im)^8} \oint\frac{d^8t}{t_{12} t_{13}t_{17}t_{18}\,t_{23}t_{24}t_{26}t_{27}}\ 
	\delta^{4|4}\!\left({\bf C}^1\cdot{\bf W}\right)\, \delta^{4|4}\!\left({\bf C}^2\cdot{\bf W}\right)\ ,
\label{G4mbox}
\ee
where
\be
	\begin{pmatrix}
		{\bf C}^1\\
		{\bf C}^2
	\end{pmatrix}
	=
	\begin{pmatrix}
		1 & t_{12} & t_{13} & t_{14}\frac{t_{23}}{t_{24}} & 0 & 0 & t_{17} & t_{18} \\
		0 & 0 & t_{23} & t_{24} & 1 & t_{26} & t_{27} & t_{27}\frac{t_{18}}{t_{17}}
	\end{pmatrix}\ .
\ee
As before, we introduce two supertwistors $\cA=(A,\chi^A)$ and $\cB=(B,\chi^B)$ by
\be
	\cA \equiv \cW^3 + \frac{t_{24}}{t_{23}}\cW^4\qquad\hbox{and}\qquad\cB \equiv \cW^7+\frac{t_{18}}{t_{17}}\cW^8
\label{4mbABdef}
\ee
so that the $\delta$-functions fix the remaining variables to be
\be
\begin{aligned}
	t_{12} = \frac{\epsilon(A,7,8,1)}{\epsilon(2,A,7,8)}\,,\quad &t_{13}=\frac{\epsilon(7,8,1,2)}{\epsilon(2,A,7,8)}\,,\quad
	&t_{17} = \frac{\epsilon(8,1,2,A)}{\epsilon(2,A,7,8)}\,,\quad &t_{18} =\frac{\epsilon(1,2,A,7)}{\epsilon(2,A,7,8)}\\
	t_{23} = \frac{\epsilon(4,5,6,B)}{\epsilon(6,B,3,4)}\,,\quad &t_{24} = \frac{\epsilon(5,6,B,3)}{\epsilon(6,B,3,4)}\,,\quad
	&t_{26} = \frac{\epsilon(B,3,4,5)}{\epsilon(6,B,3,4)}\,,\quad &t_{27} = \frac{\epsilon(3,4,5,6)}{\epsilon(6,B,3,4)}
\end{aligned}
\label{tsols}
\ee
at the expense of a Jacobian factor $[\epsilon(2,A,7,8)\epsilon(6,B,3,4)]^{-1}$. Combining all the pieces, the integral~\eqref{G4mbox} becomes
\be
	\sum R(7,8,1,2,\cA)R(3,4,5,6,\cB)\ ,
\ee
where the sum is over the two solutions for $\{\cA,\cB\}$ that we find below.

Let us see how this relates to the known formula in figure~\ref{4mbox}. We first show that $[A\wedge B]$ corresponds to a solution of the quadruple cut equations~\eqref{quadcut}. For this to be true, we must have
\be
	\epsilon(A,B,1,2) = 0\ ,\quad \epsilon(A,B,3,4) = 0\ ,\quad \epsilon(A,B,5,6) = 0\ , \quad \epsilon(A,B,7,8)=0\
\label{ABquadsol}
\ee
so that $[A\wedge B]$ intersects the four  lines $[1\wedge 2]$, $[3\wedge 4]$, $[5\wedge 6]$ and $[7\wedge 8]$ corresponding to $x_2$, $x_4$, $x_6$ and $x_8$. The second and fourth equalities in~\eqref{ABquadsol} follow because $A$ is a linear combination of $W^3$ \& $W^4$, and $B$ is likewise a linear combination of $W^7$ \& $W^8$. The first two follow upon using the ratios $t_{26}/t_{25}$ and $t_{17}/t_{18}$ given in~\eqref{tsols}. Hence $[A\wedge B]$ indeed corresponds to $x_0$ on a quadruple cut. 

In fact, one can say more. The displacement formula~\eqref{displacement} shows that
\be
\begin{aligned}
	\ell_2 \equiv x_0-x_4 &= I_{\alpha\beta}I^{\gamma\delta}
	\frac{\epsilon^\beta(\,\cdot\,,3,4,A)B_\delta-\epsilon^\beta(\,\cdot\,,3,4,B)A_\delta}{\la 34\ra\la AB\ra}\\
	&=I^{\gamma\delta}A_\delta\left(\frac{\mu^3\la 4B\ra +\mu^4\la B3\ra + \mu^B\la34\ra}{\la 34\ra\la AB\ra}\right)
\end{aligned}
\ee
so that the unprimed spinor part of $\ell_2$ (evaluated on the quadruple cut) is just the unprimed spinor part of $A$. Likewise, one can show that the unprimed spinor part of $\ell_4\equiv x_0-x_8$ is just $I^{\alpha\beta}B_\beta$. Writing
\be
	\cA = \cW^3 + a\cW^4\qquad\hbox{and}\qquad \cB= \cW^7+b\cW^8\
\ee
and substituting into the cut equations $\epsilon(A,B,1,2)=0$ and $\epsilon(A,B,5,6)=0$ shows that the ratio $a=t_{24}/t_{23}$ solves the quadratic equation $\alpha a^2 + \beta a + \gamma =0$ with coefficients
\be
\begin{aligned}
	\alpha &=\epsilon(4,5,6,\left[8)\epsilon(7\right]\!,1,2,4)\\
	\beta &= \epsilon(3,5,6,\left[8)\epsilon(7\right]\!,1,2,4)+\epsilon(4,5,6,\left[8)\epsilon(7\right]\!,1,2,3)\\
	\gamma &= \epsilon(3,5,6,\left[8)\epsilon(7\right]\!,1,2,3)
\label{quada}
\end{aligned}
\ee
while
\be
	\frac{t_{18}}{t_{17}} = \frac{\epsilon(7,1,2,3)+a\epsilon(7,1,2,4)}{\epsilon(1,2,3,8)+a\epsilon(1,2,4,8)}\ .
\ee
We have checked that the discriminant of the quadratic~\eqref{quada} is proportional to $J$.

We have therefore shown that, with the choice of the cycle in
G$(2,8)$ corresponding to the vanishing of the even cyclic Pl{\" u}cker
coordinates, we obtain a simple expression that has the right
properties to be the four-mass box coefficient: it is manifestly dual
conformal invariant and depends on the external momenta through $\sqrt J$.
Although we have not proved the full validity of~\eqref{ABquadsol}, we also remark that performing
the calculation in a different coordinate patch leads to the formula
\be
        \sum R(X,3,4,7,8)R(Y,1,2,5,6)\ \times\ 
        \frac{\epsilon(1,2,5,6)\epsilon(3,4,7,8)}{\epsilon(1,2,5,6)\epsilon(3,4,7,8)-\epsilon(3,4,X,7)\epsilon(6,Y,1,2)}\ ,
\label{XYquadsol}
\ee
where
\be
        X\equiv \cW^5 + \frac{t_{26}}{t_{25}}\cW^6\qquad\hbox{and}\qquad Y\equiv \cW^8 + \frac{t_{17}}{t_{18}}\cW^7\ .
\label{XYdef}
\ee
with $t_{26}/t_{25}$ and $t_{17}/t_{18}$ determined by substitution into
\be
\begin{aligned}
        t_{13} = \frac{\epsilon(4,X,7,8)}{\epsilon(X,7,8,3)}\ ,\quad &t_{15} = \frac{\epsilon(7,8,3,4)}{\epsilon(X,7,8,3)}\ ,\quad
        &t_{17} = \frac{\epsilon(8,3,4,X)}{\epsilon(X,7,8,3)}\ ,\quad &t_{18} = \frac{\epsilon(3,4,X,7)}{\epsilon(X,7,8,3)}\ ,\\
        t_{22} = \frac{\epsilon(5,6,Y,1)}{\epsilon(2,5,6,Y)}\ ,\quad &t_{25} = \frac{\epsilon(6,Y,1,2)}{\epsilon(2,5,6,Y)}\ ,\quad
        &t_{26} = \frac{\epsilon(Y,1,2,5)}{\epsilon(2,5,6,Y)}\ ,\quad
        &t_{28} = \frac{\epsilon(1,2,5,6)}{\epsilon(2,5,6,Y)} \, .
\end{aligned}
\label{tsolsXY}
\ee 
Since~\eqref{ABquadsol} and~\eqref{XYquadsol} are related by a GL(2) transformation of the Grassmannian, they must be equivalent.  We can identify 
\be
	R(X,3,4,7,8)R(Y,1,2,5,6) = \hat{R}_{\ell_3;648}\hat{R}_{\ell_4;826} 
\ee 
which are the same $\hat R$ invariants as in figure~\ref{4mbox}.  This nevertheless demonstrates that there are many distinct expressions for the same quantity and that our formul\ae\ have many of the required properties.


\subsection{All-loop information}
\label{sec:allloops}

In~\cite{ArkaniHamed:2009dn}, Arkani-Hamed {\it et al.} made the bold conjecture that the Grassmannian generating function really probes \emph{all-loop} information. This claim was based on the fact that in general there are more inequivalent, non-trivial contour choices than are required to reproduce the tree amplitude and 1-loop box coefficients. For example, even at NMHV level (where there are no composite residues) there are
\[
	\begin{pmatrix} n \\ 5\end{pmatrix}\ \hbox{contour choices},
\]
but only 
\[
	\frac{n(n-5)(n-6)}{2}+\frac{n(n-5)}{2} = \frac{n(n-4)(n-5)}{2}\quad \hbox{3-mass + 2-mass-hard box functions},
\]
in terms of which all the other NMHV box coefficients may be determined~\cite{Bern:2004bt}. Thus, for $n\geq8$, the Grassmannian formula contains further information, which~\cite{ArkaniHamed:2009dn} conjectured should be identified with leading singularities~\cite{Buchbinder:2005wp,Cachazo:2008dx,Cachazo:2008vp,Cachazo:2008hp,Spradlin:2008uu} of higher-loop processes. Conversely, when $n\leq7$ every choice of contour corresponds to some combination of 1-loop box coefficients, and~\cite{ArkaniHamed:2009dn} then conjectured that no new conformal invariants would be required to describe higher-loop leading singularities of NMHV amplitudes with $n\leq7$.

We do not explore this fascinating conjecture further here, but merely point out that every contour choice made in~\cite{ArkaniHamed:2009dn} may also be made here, again leading to information that is not required at 1-loop. The close correspondence of the twistor and momentum twistor Grassmannian formul\ae\ -- particularly the fact that they both probe residues on isomorphic subGrassmannians G$(k,n-4)$ -- suggests that for every possible choice of contour (after a cyclic shift in the minors) they both compute the same object. The two Grassmannian formul\ae\ then make both superconformal\footnote{As mentioned in section~\ref{sec:realtwistors}, the necessity of using (2,2) signature space-time to implement Witten's half-Fourier transform somewhat clouds the issue of usual superconformal invariance in twistor space. See also~\cite{Bargheer:2009qu,Sever:2009aa}, where the full scattering operator (rather than $n$-particle components in a Fock basis) is shown to be superconformally invariant.} and dual superconformal symmetry manifest, together with the dihedral symmetry of the colour-ordered amplitudes.


\subsection{$\overline{\rm MHV}$ amplitudes}
\label{sec:MHVbaramps}

For $\overline{\rm MHV}$ amplitudes ($k=n-4$), the $\delta$-functions saturate the integral and no extra contour need be specified. However, since we are working in a chiral framework, the $\overline{\rm MHV}$ amplitudes are not trivial in momentum twistor space (unlike the MHV amplitudes). Using the coordinate patch 
\be
	\begin{pmatrix}
		{\bf T}^1\\ \vdots \\{\bf T}^{n-4}
	\end{pmatrix}
	=
	\begin{pmatrix}
		1 & \cdots & 0 & t^1_{\ n-3} & t^1_{\ n-2} & t^1_{\ n-1} & t^1_{\ n}\\
		\vdots &\ddots & \vdots & \vdots & & & \vdots\\
		0 & \cdots & 1 & t^k_{\ n-3} & t^k_{\ n-2} & t^k_{\ n-1} & t^k_{\ n}
	\end{pmatrix}
\ee	
the $\delta$-functions enforce
\be
\begin{aligned}
	t^r_{\ n-3} &= \frac{\epsilon(n\!-\!2,n\!-\!1,n,r)}{\epsilon(n\!-\!3,n\!-\!2,n\!-\!1,n)} 
	\qquad 
	&t^r_{\ n-2} &= \frac{\epsilon(r,n\!-\!3,n\!-\!1,n)}{\epsilon(n\!-\!3,n\!-\!2,n\!-\!1,n)}\\
	t^r_{\ n-1} &= \frac{\epsilon(n\!-\!3,n\!-\!2,n,r)}{\epsilon(n\!-\!3,n\!-\!2,n\!-\!1,n)}
	\qquad
	&t^r_{\ n}   &= \frac{\epsilon(r,n\!-\!3,n\!-\!2,n\!-\!1)}{\epsilon(n\!-\!3,n\!-\!2,n\!-\!1,n)}
\end{aligned}
\ee
for $r=1,\ldots,k=n-4$. Plugging these values into the cyclic minors, one finds after some algebra that 
\be
	\frac{A_{n,0}^{\overline{\rm MHV}}}{A_{n,0}^{\rm MHV}} =
	\prod_{k=1}^{n}\frac{1}{\epsilon(k,k\!+\!1,k\!+\!2,k\!+\!3)}\ 
	\frac{\prod_{i=1}^{n-4}\delta^{0|4}(\chi^i\epsilon(i\!+\!1,i\!+\!2,i\!+\!3,i\!+\!4)+\hbox{cyclic})}
	{\prod_{j=2}^{n-4}\epsilon(j,j\!+\!1,j\!+\!2,j\!+\!3)^4}
\label{MHVbar}
\ee
in momentum twistor space. The first product is manifestly cyclic. The
remaining factors can also be shown to be cyclically invariant, with
the denominator acting as a Jacobian to compensate for the cyclic
shift of the fermionic $\delta$-functions in the numerator. We have
checked for $n\leq7$ that~\eqref{MHVbar} agrees with the momentum
space expression given in~\cite{Drummond:2008cr}. 


\section{Polytopes}
\label{sec:polytopes}

The dual superconformal invariants relevant for NMHV tree amplitudes have been previously studied in momentum twistor space by Hodges~\cite{Hodges:2009hk}. In this section we make a formal connection between our Grassmannian approach and ref.~\cite{Hodges:2009hk}. We first give a brief review of Hodges' essential ideas.  In his picture, each basic invariant
$R(a,b,c,d,e)$ is interpreted as the `holomorphic volume' of a certain
4-simplex in\footnote{With Penrose conventions, the twistor space with
  coordinates $Z^\alpha$ would usually be taken as primary, and the
  $W_\alpha$ space referred to as `dual'.  This unfortunately clashes
  with the prevalent conventions in perturbative gauge theory, whereby
  MHV amplitudes involve unprimed/undotted spinors $|\lambda\ra$ and
  so live most naturally on Penrose's dual space.}  \emph{dual}
momentum twistor space with coordinates $\cZ^I$: 
\be
\label{simplex}
	R(1,2,3,4,5) = \int_{\rm simplex} \hspace{-0.5cm}d^{4|4}\cZ\ .  
\ee 
Here, the integral is a real $4|4$-dimensional integral over a contour in
$\mathbb{C}^{4|4}$, where the contour has boundary on the simplex whose facets (codimension-one
faces) are the planes 
\be 
	\cZ^I\cW_I^j = 0 \qquad\hbox{for}\quad j=1,\ldots,5
\label{Hsimplex}
\ee
determined by the external twistors $\cW^i$.

\begin{figure}[t]
\begin{center}
	\includegraphics[height=45mm]{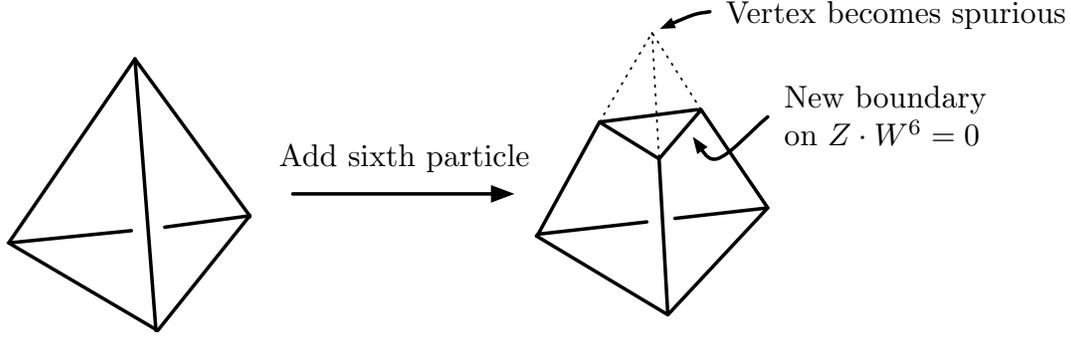}
\end{center}
\caption{{\it Adding a new particle truncates the previous polytope along a new plane, shown here for $5\to6$ particles, projected into the plane $Z\cdot W^5=0$. The vertex where planes $Z\cdot W^5=0$, $Z\cdot W^1=0$, $Z\cdot W^2=0$ and $Z\cdot W^3=0$ meet becomes spurious when  particle 6 is added, and the volume of the resulting polytope stays finite even when this vertex moves to infinity}.  }
\label{fig:polytopes}
\end{figure}

The power of this interpretation is that one can understand why it is
natural to consider \emph{sums} of $R$-invariants \emph{with equal
  coefficients}, such as arise in the NMHV tree amplitude: the sum is
simply the (oriented) volume of the polytope made up from the union of
the elementary simplices (with appropriate signs). For example, the BCFW sum 
\be 
	\frac{A_{6,0}^{\rm NMHV}}{A_{6,0}^{\rm MHV}} = R(1,2,3,4,5) - R(3,4,5,6,1) + R(5,6,1,2,3)
\label{BCFW6}
\ee
corresponds to the volume of a 6-sided polytope with dihedral symmetry.  The BCFW representation is obtained by dividing up the polytope into elementary simplices $R(a,b,c,d,e)$. However, such decompositions are not unique, and the decomposition obtained by a cyclic shift of $(1,2,3,4,5,6)$ yields the alternative BCFW formula
\be
	\frac{A_{6,0}^{\rm NMHV}}{A_{6,0}^{\rm MHV}} = -R(2,3,4,5,6) + R(4,5,6,1,2)-R(6,1,2,3,4)
\ee
for the same volume. In the case of the split helicity amplitude, one of the terms vanishes and we
can reduce to a 3-dimensional picture in which the new
bounding plane slices off one of the vertices (see figure~\ref{fig:polytopes}).  

The spurious singularities inherent in the BCFW decomposition of an amplitude may be
understood from this picture. The polytope's volume diverges if any of its
vertices move away to infinity -- this occurs when any four twistors $W^i$ become coplanar, so that the four corresponding facets meet at infinity.  Such a singularity is physical when the four twistors form two consecutive pairs; if the pairs are $\{W^{i-1},W^i\}$ and $\{W^{j-1},W^j\}$ the pole corresponds to the singularity
\be
	(p_{i}+\ldots +p_{j})^2\to 0
\ee
of the amplitude. By construction, all the vertices of the full polytope (characterising the overall amplitude) are of this form. However, any BCFW decomposition also contains spurious singularities, corresponding to vertices where four `non-pairwise consecutive' facets meet -- in slicing up the polytope to obtain a BCFW decomposition, one necessarily introduces new  vertices, that may be either internal or external to the full polytope. The volumes of the individual elementary simplices of the decomposition depend on the positions of these spurious vertices, and accordingly they have spurious singularities as these vertices move away to infinity  (see figure~\ref{fig:polytopes}).

\bigskip

We would now like to understand the relation of Hodges' polytopes to
the Grassmannian contour integral 
\be 
	R_{k,n} = \oint_{\Gamma\subset{\rm G}(k,n)} \hspace{-0.5cm}
	d\mu\hspace{0.5cm} \prod_{r=1}^k\delta^{4|4}({\bf T}^r{\bf \cdot W}) 
\ee 
We first discuss the basic 5-twistor $R$-invariant, making
contact with the real, split signature version of the Grassmannian formula~\eqref{Rcont-split} in a very formal fashion.  In this case we can take the $\cW_i$ to be real and the contour is required to lie in the real twistor space.  To incorporate the contour into the integral, we introduce step functions 
\be
\label{step-fn}
	\theta(x)\equiv\frac1{2\pi \im}\int \mathrm{e}^{-\im Tx}\frac{ dT}{T+\im\varepsilon}
	= \begin{cases}
		1 \quad \hbox{for}\ x > 0\cr 
		\frac{1}{2} \quad \hbox{for} \ x=0
		\cr 0 \quad \hbox{for}\ x< 0 
	\end{cases}\ .
\ee 
This can be used formally to determine the simplex by writing our integral as
\be
\begin{aligned}
	\int_{\mathrm{simplex}} \hspace{-0.5cm}d^{4|4} \cZ\ 
	&= \int d^{4|4}\cZ  \ \prod_{i=1}^5\theta(\cW^i\cdot\cZ) \\
	&=  \int d^{4|4}\cZ \ \prod_{i=1}^5{\rm e}^{-\im T_i\cW^i\cdot\cZ}\frac{d T_i}{T_i+\im\varepsilon} \\
	&=  \int \prod_{i=1}^5\frac{d T_i}{T_i+\im\varepsilon}\ \delta^{4|4}\!\left(\sum_{i=1}^5 T_i\cW^i\right) 
\label{hodge-der}
\end{aligned}
\ee
where in the second line we have simply substituted in the definition
of the step functions~\eqref{step-fn} and in the last line we have integrated out $\cZ$, obtaining the $\delta$-functions.  This has one more integral than there are bosonic $\delta$-functions. The homogeneity of the $\delta$-function allows us to factor out an overall scale, say $T_1$. Writing $t_i=T_i/T_1$ for $i=2,\ldots ,5$ we can rewrite the last integral as 
\be
	\int  \frac{d T_1}{T_1+\im\varepsilon}\ \prod_{i=2}^5\frac{d t_i}{t_i+\im\varepsilon/T_1}
	\ \delta^{4|4}\!\left(\cW_1+\sum_{i=2}^5 t_i\cW_i\right)  \, .
\ee
Since the $t_i$ are now to be integrated against $\delta$-functions, the
regularization $+\im\varepsilon/T_1$ is immaterial -- in particular its $T_1$ dependence
can be ignored.  The $T_1$ integral can therefore be performed directly, giving an overall factor of 
$1/2$ and we obtain the formula for the dual conformal invariant~\eqref{Rcont-split}, up to an overall sign.

This calculation is formal in two related respects.  Firstly,
we  have not obtained the required sign factor ${\rm sgn}(\epsilon(2,3,4,5))$ and secondly, the region of 
integration in~\eqref{hodge-der} depends on the signs of the $\cW^i$. Let us give a brief sketch of the
ideas required to put this calculation on a firmer footing.

Hodges' contour --  the simplex in~\eqref{simplex} -- can be parametrized
explicitly using real bosonic parameters $\lambda_i$ (with $i=1,\ldots 5$) and real fermionic parameters $\psi$ as 
\be
	\cZ=\sum_{i=1}^5 \lambda_i (V_i,\psi) \,, \quad \mbox{where}\quad 
	\lambda_i >0 \quad \hbox{and}\quad \sum\lambda_i=1
\ee
and $\cV_i=(V^{\alpha}_i,\psi^a_i)$ are the coordinates of the vertices of the simplex. The bosonic vertex coordinates $V_i$ are determined in terms of $\psi$ and the external twistors $\cW^j$ by solving the four equations 
\be
	0=\cV_i\cdot\cW^j=V^{\alpha}_iW^j_\alpha+\psi^a \chi_a^j\qquad\hbox{when}\qquad i\neq j\ .
\ee
For example,
\be
	V^\alpha_1= -\frac{\psi^a}{\epsilon(2,3,4,5)}\left(\chi_a^2 \epsilon^\alpha(\,\cdot\,,3,4,5) + \mbox{cyclic}\right)\ .
\ee
while the remaining $V_i$ are related to this by cyclic permutations.  Thus we can write
\be
	\lambda_i= \frac{\cZ\cdot\cW^i}{\cV_i\cdot\cW^i}
\ee
(where the dot product indicates contraction over the supertwistor indices). Therefore the appropriate integral for the supersymmetric volume of the simplex is 
\be
	\int d^{4|4}\cZ \ \prod_{i=1}^5 \theta(\lambda_i)
\ee
This resolves the dependence on the signs of $\cW^i$ (the $\lambda_i$
are weightless in  $\cW^i$) and introduces the additional sign
factors in the final formula.  The ideal derivation would lead
directly to the holomorphic formula~\eqref{Rcont}, either via contour
integrals or the Dolbeault form for $R_{1,5}$.

For the polytopes appropriate to NMHV amplitudes with more than 5
points, we simply introduce additional factors of $\theta(\lambda_i)$, one for each new particle. 
In the formal calculation it is clear that this leads to the expression
\be
	\int \prod_{i=1}^n\frac{d T_i}{T_i+\im\varepsilon}\ \delta^{4|4}\!\left({\bf T\cdot \cW}\right)\ .
\ee
There are $n-4$ more integrals than $\delta$-functions, so these
integrals must be done in the $T_i$ parameter space (the
non-projective Grassmannian) as the delta function parts of the
$1/t+i\epsilon$ distribution.

We have therefore seen that in formal terms, the relationship between
our intgrals over projective space for NMHV amplitudes and Hodges
volumes of polytopes in dual momentum twistor space can simply be
understood via the Fourier transform from momentum twistor space to
dual momentum twistor space in a totally real formulation.  For
N$^{k}$MHV, it is clear that this idea can be extended by expressing
each delta function as an integral over dual momentum twistors space
leading to an integral over the $(k-2)$-fold cartesian product of such
dual momentum twistor spaces.  The conversion of the $T$-integrals
into step functions on this space is no longer so clear because the
denominators are no longer simple factors, so the geometric
interprestation is less clear.


\section{Discussion}
\label{sec:conclusions}

In this paper, we have translated the various dual conformal
invariants into momentum twistor space -- the twistor space of region
momenta. We have shown that these invariants are naturally
generated by the remarkable contour integral
\be
	\frac{1}{(2\pi\im)^{k(n-k)}}\oint d\mu\  \prod_{r=1}^k\, \delta^{4|4}\!\left({\bf T}^r\cdot{\bf W}\right)\ ,
\label{Glast}
\ee
where, after performing the integrals over the $\delta$-functions, the
contour is taken to be a suitable cycle in G$(k,n-4)$. This formula
makes manifest both dual superconformal invariance and (at least for
appropriate choices of contour) the dihedral symmetry of the
colour-ordered amplitude. More importantly, the possibility of
enclosing several poles with the same contour allows one to take sums
of dual superconformal invariants \emph{with equal coefficients} (up
to sign). Similarly, although we have not emphasised this point in the
present paper, higher-dimensional versions of Cauchy's theorem provide
a natural way to understand the many identities that these sums of
invariants obey -- identities that become ever more algebraically
involved as $(n,k)$ increase. A loose end in our discussion is the identification of the N$^2$MHV
4-mass box coefficient, where we have not yet been able to prove that our formula agrees with the formula in~\cite{Drummond:2008bq,Hall:2009xg}.  

There are many other questions that this work leaves open. Firstly,
our Grassmannian
formula \eqref{Glast} is clearly analogous to the Grassmannian generating
principle found by Arkani-Hamed {\it et al.}
in~\cite{ArkaniHamed:2009dn} that naturally lives in ordinary twistor
space. As we have seen, this analogy extends even to the
specification of the appropriate contours at the zero sets of the
cyclic Pl{\" u}cker coordinates.  This obviously deserves a more
direct understanding. Clearly, one can mechanically translate
between amplitudes on ordinary twistor space and on momentum
twistor space by first using Witten's half-Fourier
transform~\cite{Witten:2003nn} and then performing the algebraic change of variables we have explored in this paper. However, we feel that there should be a more direct relationship between the two twistor spaces, that does not require translating via momentum space. Could it be that the usual and momentum twistor
spaces are T-dual?  We have
recently heard\footnote{Arkani-Hamed \& Cachazo, private communication}
that it is possible to demonstrate a direct correspondence between the
integrals over cycles in these two distinct Grassmannians.
(In particular, this correspondence demonstrates the validity of our N$^2$MHV
box coefficient).

The present paper has been concerned with tree amplitudes and leading
singularities of loop amplitudes. However, it is  
natural to wonder whether it is possible to `dress' (either of the)
Grassmannian integrals so as to obtain the loop amplitudes proper. In
this regard, we would like to point out that it is straightforward to
translate the scalar box integrals to momentum twistor space. For
example, a generic 1-loop scalar box integral may be regularized by
the Feynman $\im\varepsilon$-prescription\footnote{We particularly
  thank Andrew Hodges and James Drummond for discussions of this
  point.} 
\be
I_{rstu}^{\varepsilon}= \int \frac{d^4x_0}
{(x_{0r}^2+\im\varepsilon)(x_{0s}^2+\im\varepsilon)(x_{0t}^2+\im\varepsilon)(x_{0u}^2+\im\varepsilon)}\ ,
\label{scalarint}
\ee
with $\lim_{\varepsilon\to0}$ finite only for the four-mass box. The standard form~\eqref{twistdist} of the metric in twistor variables shows that this integral may be written as
\be
	\oint  \frac{D^3U\wedge D^3V\ \la r\!-\!1\ r\ra\la s\!-\!1\ s\ra\la t\!-\!1\ t\ra\la u\!-\!1\ u\ra}
	{\left(\epsilon(U,V,r\!-\!1,r)+\im\varepsilon\la UV\ra\la r\!-\!1\  r\ra\right)\cdots
	\left(\epsilon(U,V,u\!-\!1,u)+\im\varepsilon\la UV\ra\la u\!-\!1\ u\ra\right)}
\label{scalarinttwist}
\ee
in momentum twistor space, where $[U\wedge V]$ is the line in twistor
space corresponding to the point $x_0$, $D^3U$ is the canonical
holomorphic measure of weight $+4$ on $\CP^3$ and the contour is the
diagonal $\Delta\subset\CP^3\times\CP^3$. The challenge here is to
combine~\eqref{scalarinttwist} with~\eqref{Glast} in a way that does
not ruin all the beautiful properties~\eqref{Glast} possesses. In
particular, one would \emph{not} wish to follow the usual line of
simply multiplying a box coefficient (obtained by imposing some
particular contour on~\eqref{Glast}) by its corresponding box function
(written as in~\eqref{scalarinttwist}), and then summing over
boxes. Such an approach undoes one of the main benefits of the
Grassmannian -- that identities such as mysterious combinations of box
functions being IR finite can be understood~\cite{ArkaniHamed:2009dn}
via the global residue theorem! Presumably, a successful unification
will involve thinking of the box coefficients as leading singularities
of a true loop amplitude; that is, the leading singularity is an
evaluation of the usual loop amplitude over a $T^{4\ell}$ contour that
encircles the poles of the propagators (and `hidden
propagators'~\cite{Buchbinder:2005wp,Cachazo:2008dx}) in the space
$\C^{4\ell}$ of complex momenta, rather than 
the usual $\mathbb{R}^{4\ell}$ contour. Similarly, the Grassmannian
generating function~\eqref{Glast} should itself emerge as the leading
singularity of some larger object that depends on some internal
twistors and knows about the full amplitude.

In section~\ref{sec:polytopes} we used real methods such as the
Fourier transform to express the $\delta$-functions as integrals over
regions in the dual momentum twistor space. This led us to
Hodges' picture~\cite{Hodges:2009hk} of amplitudes as volumes of polytopes. Our investigation used a representation of the dual conformal invariants in \emph{real} twistor space; it would be useful to have a proof that is truer to the more appropriate \emph{holomorphic} objects used in the rest of this
paper. Such a proof should be based on contour integrals and
the twistor transform rather than the Fourier transform.  However,
this requires a better understanding how to use twistor elemental states, introduced in~\cite{Mason:2009sa} in the complex setting. Nevertheless, this real approach is sufficient for us to see that the
extension of Hodges' approach to N$^k$MHV will not have such a simple interpretation as for NMHV
since, although the same idea can be used to express terms in the BCFW decomposition of an N$^k$MHV amplitude as
integrals over the $(k-2)$-fold product of dual twistor space, the
integrand will no longer be expressible as a straightforward product of step functions.  It will be interesting to see whether one can nevertheless turn this picture into a useful
formulation.  Certainly, Hodges' original formulation for NMHV
amplitudes gave a beautiful geometric intepretation of the full
NMHV amplitude that promises more for the full amplitude.

\vspace{2cm}
\noindent{\Large\bf Acknowledgments}\\

\noindent We would like to thank Freddy Cachazo, Cliff Cheung, James
Drummond, Andrew Hodges, Jared Kaplan and especially Nima Arkani-Hamed
for many useful discussions. We would also like to thank the
organizers of the ``Integrability in Gauge and String Theory''
conference, and the faculty \& staff of AEI, Golm for a very enjoyable
stay. Finally, we thank the referee for a number of useful comments that helped to improve the manuscript.
 The work of DS was supported by the Perimeter Institute for
Theoretical Physics. Research at the Perimeter Institute is supported
by the Government of Canada through Industry Canada and by the
Province of Ontario through the Ministry of Research \&
Innovation. This work was financed by EPSRC grant number EP/F016654,
see also {\tt
  http://gow.epsrc.ac.uk/ViewGrant.aspx?GrantRef=EP/F016654/1}.

\appendix

\section{Higher-order invariants}
\label{app:higherorder}

In this appendix we translate the higher-order dual conformal invariants $R_{n;a_1b_1;a_2b_2;\ldots;a_rb_r;ab}$ into momentum twistor space. These play a role in N$^{r+1}$MHV amplitudes for $r\geq1$ and were defined in~\cite{Drummond:2008cr} by\footnote{Following~\cite{Drummond:2009ge}, we have slightly rearranged the indices  of these invariants compared to their original definition in~\cite{Drummond:2008cr}, so that $R_{n;ab}$ is naturally the $r=0$ case of the general structure.}
\be
	R_{n;a_1b_1;a_2b_2;\ldots;a_rb_r;ab}
	\equiv
	\frac{\la a\ a\!-\!1\ra\la b\ b\!-\!1\ra\ \delta^{0|4}\!\left(\la\xi|x_{a_ra}x_{ab}|\theta_{ba_r}\ra
	+\la\xi|x_{a_rb}x_{ba}|\theta_{aa_r}\ra\right)}{x_{ab}^2\la\xi|x_{a_rb}x_{ba}|a\!-\!1\ra\la\xi|x_{a_rb}x_{ba}|a\ra
	\la\xi|x_{a_ra}x_{ab}|b\!-\!1\ra\la\xi|x_{a_ra}x_{ab}|b\ra}
\label{higherorder}
\ee
where
\be
	\la\xi|\equiv\la n|x_{nb_1}x_{b_1a_1}x_{a_1b_2}\cdots x_{b_ra_r}\ .
\ee
Using~\eqref{displacement}, it is straightforward to see that $\la\xi|=I^{\alpha\beta}U_\beta$, where
\be
\begin{aligned}
	U_{\sigma}&\equiv n_\alpha
	\left(\frac{[n\!-\!1\wedge n]}{\la n\!-\!1\ n\ra}
	-\frac{[b_1\!-\!1\wedge b_1]}{\la b_1\!-\!1\ b_1\ra}\right)^{\alpha\beta}
	\left(\frac{[b_1\!-\!1\wedge b_1]}{\la b_1\!-\!1\ b_1\ra}
	-\frac{[a_1\!-\!1\wedge a_1]}{\la a_1\!-\!1\  a_1\ra}\right)_{\beta\gamma}\\
	&\qquad\qquad \times \left(\frac{[a_1\!-\!1\wedge a_1]}{\la a_1\!-\!1\  a_1\ra}
	-\frac{[b_2\!-\!1\wedge b_2]}{\la b_2\!-\!1\ b_2\ra}\right)^{\gamma\delta}
	\quad\cdots\quad 
	\left(\frac{[b_r\!-\!1\wedge b_r]}{\la b_r\!-\!1\ b_r\ra}-\frac{[a_r\!-\!1\wedge a_r]}{\la a_r\!-\!1\ a_r\ra}\right)_{\rho\sigma}\\
	&=\frac{\epsilon(n,b_1\!-\!1,b_1,[a_1\!-\!1)\epsilon(a_1],b_2\!-\!1,b_2,[a_2\!-\!1)
	\ \cdots\ \epsilon(a_{r-1}],b_r\!-\!1,b_r,[a_r\!-\!1),a_r]_{\sigma}}
	{\la b_1\!-\!1\ b_1\ra\la a_1\!-\!1\ a_1\ra\la b_2\!-\!1\ b_2\ra\la a_2\!-\!1\ a_2\ra
	\ \cdots\ \la b_r\!-\!1\ b_r\ra\la a_r\!-\!1\ a_r\ra}
\end{aligned}
\label{appUdef}
\ee
where in the first line, indices are raised with the totally-skew $\epsilon$-symbol and $[n\!-\!1\wedge n]_{\alpha\beta}$ denotes the skew twistor $W^{n-1}_{[\alpha}W^n_{\beta]}$ (the line through twistors $n\!-\!1$ and $n$), while in the second line, the square brackets again denote antisymmetrization. The denominator may now be translated exactly as for the basic invariant, with the replacement $W^n\to U$. 

Similarly, we can use the identity
\be
	\la\xi|x_{a_ra}x_{ab}|\theta_{ba_r}\ra+\la\xi|x_{a_rb}x_{ba}|\theta_{aa_r}\ra
	=x_{ab}^2\la\xi|\theta_{a_r}\ra + \la\xi|x_{a_ra}x_{ab}|\theta_b\ra +\la\xi|x_{a_rb}x_{ba}|\theta_{a}\ra
\ee
to translate the numerator. Just as in~\eqref{fermitrans2}, the final two terms become
\be
	\la\xi|x_{a_ra}x_{ab}|\theta_b\ra \to 
	\frac{\epsilon(U,a\!-\!1,a,b\!-\!1)\chi^b-\epsilon(U,a\!-\!1,a,b)\chi^{b-1}}{\la a\!-\!1\ a\ra\la b\!-\!1\ b\ra}
\ee
and similarly for $\la\xi|x_{a_rb}x_{ba}|\theta_a\ra$. The remaining term involves
\be
	\la\xi|\theta_{a_r}\ra =  I^{\rho\sigma}U_\rho 
	\left(\frac{\chi^{a_r-1}W_\sigma^{a_r} - \chi^{a_r}W_{\sigma}^{a_r-1}}{\la a_r\!-\!1\ a_r\ra}\right)\ .
\ee
Equation~\eqref{appUdef} shows that $U_\rho = =\alpha W^{a_r}_\rho - \beta W^{a_r-1}_\rho$, so this is simply
\be
	 \chi^U\equiv\left(\alpha \chi^{a_r-1}- \beta\chi^{a_r}\right)\ ,
\ee
where the definition of $\chi^U$ is motivated by the fact that it has exactly the same form~\eqref{appUdef} as the bosonic twistor $U_\alpha$, but with the free index being the fermionic part of the final supertwistor. We thus find
\be
	x_{ab}^2\la\xi|\theta_{a_r}\ra = \frac{\epsilon(a\!-\!1,a,b\!-\!1,b)}{\la a\!-\!1\ a\ra\la b\!-\!1\ b\ra}\chi^U\ .
\ee
It is natural to extend $U$ to a supertwistor $\cU_I = (U_{\alpha},\chi^U_a)$, whereupon the higher-order dual conformal invariants~\eqref{higherorder} take exactly the same form in momentum twistor space as the first-order $R$-invariants, except with $\cW^n\to\cU$. Explicitly,
\be
	R_{n;a_1b_1;a_2b_2;\ldots;a_rb_r;ab} = R(\cU,a\!-\!1,a,b\!-\!1,b)
\ee
as used in section~\ref{sec:NNMHVtrees}. The `boundary terms' in Drummond \& Henn's solution may be handled similarly.


\end{document}